# Ordered Counter-Abstraction[*]

Ahmed Rezine

Linköping University, Sweden

**Abstract.** We introduce a new symbolic representation based on an original generalization of counter abstraction. Unlike classical counter abstraction (used in the analysis of parameterized systems with unordered or unstructured topologies) the new representation is tailored for proving properties of linearly ordered parameterized systems, i.e., systems with arbitrary many finite processes placed in an array. The relative positions in the array capture the relative priorities of the processes. Configurations of such systems are finite words of arbitrary lengths. The processes communicate using global transitions constrained by their relative priorities. Intuitively, an element of the symbolic representation has a base and a set of counters. It denotes configurations that respect the constraints imposed by the counters and that have the base as a subword. We use the new representation in a uniform and automatic Counter Example Guided Refinement scheme. We introduce a relaxation operator that allows a well quasi ordering argument for the termination of each iteration of the refinement loop. We explain how to refine the relaxation to systematically prune out false positives. We implemented a tool to illustrate the approach on a number of parameterized systems.

## 1 Introduction

We introduce in this paper an original adaptation of counter abstraction and use it for the verification of safety properties for linearly ordered parameterized systems. Typically, such a system consists of an arbitrary number of identical processes placed in a linear array. Each process is assumed to have a finite number of states. The arbitrary size of these systems results in an infinite number of possible configurations. Examples of linearly ordered parameterized systems include mutual exclusion algorithms, bus protocols, telecommunication protocols, and cache coherence protocols. The goal is to check correctness regardless of the number of processes in the system.

Configurations of a parameterized system can be seen as finite words of arbitrary lengths over the finite set $Q$ of process states. Processes change state using transitions that might involve global conditions. These can be universal or existential. Transition (1) below is constrained by a universal condition. It requires that a process (with array index) $i$ may perform the transition only if all processes with indices $j > i$ (i.e., to the right of $i$, hence $\forall_R$) are in states $\{q_1, q_2, q_3\} \subseteq Q$.

$$t : q_5 \to q_6 : \forall_R \{q_1, q_2, q_3\} \tag{1}$$

An existential condition may require that some (instead of all) processes with indices $j > i$ are in certain states. Regular model checking [18,11] is an important technique which has been used for the uniform verification of infinite state systems in general, and of linearly ordered parameterized systems in particular. This technique uses finite state automata to represent sets of configurations, and transducers (i.e., finite state automata over pairs of letters) to capture transitions of the system. Verification boils down to the repeated calculation of several automata-based constructions among which is the application of the transducers to (typically) heavier and heavier automata representing more and more complex sets of reachable configurations. To ease termination of these computations, acceleration [5], widening [8,22] and abstraction [9] methods are used.

---

[*] Work supported in part by project 12.04 of the CENIIT research organization, Linköping.

In order to combat this complexity, the framework of monotonic abstraction [4,3] uses upward closed sets (wrt. a predefined pre-order) as symbolic representations. This introduces an over-approximation, as sets of states generated during the analysis are not necessarily upward closed. The advantage is to use minimal constraints (instead of arbitrary automata) to succinctely represent possibly infinite sets of configurations. The approach typically adopts the subword relation as pre-order for the kind of systems we consider in this work. As a concrete example, if $q_5 \in Q$, then the word $q_5 q_5$ would represent all configurations in $(Q^* q_5 Q^* q_5 Q^*)$ since $q_5 q_5$ is subword of each one of them. The analysis starts with upward closed sets and repeatedly approximates sets of predecessors by closing them upwards. Termination is guaranteed using a well quasi ordering argument [16]. The scheme proved quite successful [4,3,2] but did not propose refinements for pruning false positives for ordered systems like the ones we consider here. Resulting approximations are particularly inadequate when performing forward analysis, which can be more efficient [15] in general.

In this work, we augment precision on demand by combining the use of minimal constraints a la monotonic abstraction with threshold based counter abstraction. The idea of counter abstraction [21,7,13,17] is to keep track of the number of processes which satisfy a certain property. A typical property for a process is to be in a certain state in $Q$. A simple approach to ensure termination is then to count up to a prefixed threshold. After the threshold, any number of processes satisfying the property is assumed possible. This results in a finite state system that can easily be analyzed. If the approximation is too coarse, the threshold can be augmented. For systems like those we consider in this paper, automatically finding the right properties and thresholds can become very challenging. Consider for instance the transition (1) above. It is part of Burns mutual exclusion algorithm, where $q_6$ models access to the critical section (see appendix). Suppose we want to compute the $t$-successors of configurations only containing processes in state $q_5$. These are in fact reachable in Burns algorithm. Plain counter abstraction would capture that all processes are at state $q_5$. After one step it would capture that there is one process at state $q_6$ and all other processes are at state $q_5$ (loosing that $q_6$ is at the right of all $q_5$, if any). After the second step it would conclude that configurations with at least two $q_6$ are also reachable (thus violating mutual exclusion). Observe that augmenting a threshold would not help as the problem is inherent to the loss of information about the relative positions of the processes. Upward closure based representations will also result in a mutual exclusion violation if used in forward on this example. Suppose we use $q_5 q_5$ as a minimal constraint. Upward closure wrt. to the subword relation would result in the set $(Q^* q_5 Q^* q_5 Q^*)$ which also allows two processes at state $q_6$ to coexist. Even when using the refined ordering of [3], upward closure would result in

$(\{q_5\}^* q_5 \{q_5\}^* q_5 \{q_5\}^*)$. After one step, the obtained $(\{q_5\}^* q_5 \{q_5\}^* q_6)$ will be approximated with $(\{q_5, q_6\}^* q_5 \{q_5, q_6\}^* q_6 \{q_5, q_6\}^*)$, again violating mutual exclusion. Approximations are needed to ensure termination of the analysis. Indeed, without approximation, one would have to differentiate infinite numbers of sets, like the sequence

$$(\{q_5\}^* q_6), (\{q_5\}^* q_6 \{q_5\}^* q_6), \ldots (\{q_5\}^* q_6 \{q_5\}^* \ldots \{q_5\}^* q_6) \qquad (2)$$

The idea of this work is to combine threshold-based counter abstraction together with upward closure techniques in order to gain precision while still ensuring termination. To achieve this, we introduce the notion of a *counted word*. A counted word has a *base* and a number of formulae (called *counters*). Like in monotonic abstraction, a base (a word in $Q^*$) is used as a minimal element and denotes all larger words wrt. the subword relation. In addition, the counters are used to constrain the denotation of the base. We associate two counters per state in the base. For each state, one counter (called *left counter*) constrains Parikh images of allowed prefixes to the left of the state, and the other (called *right counter*) constrains Parikh images of allowed suffixes to the right of the state. For example $(\{q_5\}^* q_6)$, which cannot be captured by usual upward closure or counter abstraction techniques, is captured by the counted word $\varphi_1$ in the sequence:

$$\varphi_1 = \left(\begin{bmatrix} v_{q_5} \geq 0 \\ \wedge v_{q_6} = 0 \end{bmatrix}, q_6, \begin{bmatrix} v_{q_5} = 0 \\ \wedge v_{q_6} = 0 \end{bmatrix}\right)$$

$$\varphi_2 = \left(\begin{bmatrix} v_{q_5} \geq 0 \\ \wedge v_{q_6} = 0 \end{bmatrix}, q_6, \begin{bmatrix} v_{q_5} \geq 0 \\ \wedge v_{q_6} = 1 \end{bmatrix}\right)\left(\begin{bmatrix} v_{q_5} \geq 0 \\ \wedge v_{q_6} = 1 \end{bmatrix}, q_6, \begin{bmatrix} v_{q_5} = 0 \\ \wedge v_{q_6} = 0 \end{bmatrix}\right)$$

$$\vdots$$

$$\varphi_k = \left(\begin{bmatrix} v_{q_5} \geq 0 \\ \wedge v_{q_6} = 0 \end{bmatrix}, q_6, \begin{bmatrix} v_{q_5} \geq 0 \\ \wedge v_{q_6} = (k-1) \end{bmatrix}\right) \cdots \left(\begin{bmatrix} v_{q_5} \geq 0 \\ \wedge v_{q_6} = (k-1) \end{bmatrix}, q_6, \begin{bmatrix} v_{q_5} = 0 \\ \wedge v_{q_6} = 0 \end{bmatrix}\right)$$

In $\varphi_1$, the base $q_6$ denotes the set $(Q^* q_6 Q^*)$. This is constrained to $(\{q_5\}^* q_6 Q^*)$ by the right counter $\begin{bmatrix} v_{q_5} \geq 0 \\ \wedge v_{q_6} = 0 \end{bmatrix}$ and to $(\{q_5\}^* q_6)$ by the left counter $\begin{bmatrix} v_{q_5} = 0 \\ \wedge v_{q_6} = 0 \end{bmatrix}$. Sequence (2) can then be captured by the counted words $\varphi_1, \varphi_2, \ldots \varphi_i$. This gain in precision comes at the cost of termination. We therefore use a family of *relaxations*. Each relaxation comes with thresholds associated to each state in $Q$. If a counter requires $(v_q = k)$ with $k$ larger than the threshold imposed by the relaxation, we weaken $(v_q = k)$ into $(v_q \geq k)$. Using a well quasi ordering argument, we show that this is enough to ensure termination of the analysis that relaxes all manipulated counted words. If the relaxation is too coarse and generates a spurious trace, we propose a mechanism to detect states the thresholds of which need to be increased in order to get rid of the spurious trace. The scheme can be used both in forward or in backward analysis in order to check reachability of sets of configurations. We tried the approach on a prototype implementation and obtained encouraging results on a number of mutex algorithms.

*Related work.* Other verification efforts with a termination guaranty typically consider decidable subclasses [1,13,12], or use approximations to obtain systems on which the analysis is decidable [7,2] and [21,10,20]. For example, the authors in [12] propose a forward framework with systematic refinement to decide safety properties for a decidable class. The problem we consider here is undecidable. The authors in [20] use heuristics to deduce cut-offs in order to check invariants on finite instances. In [21] the authors use counter abstraction and truncate the counters in order to obtain a finite state system. This might require manual insertion of auxiliary variables to capture the relative order of processes in the array. Environment abstraction [10] combines predicate and counter abstraction. This allows it to handle systems where processes manipulate infinite variables (e.g. identifiers). It also results in what is essentially a finite state approximated system. Hence, it can require considerable interaction and human ingenuity to find the right predicates. Our approach handles linearly ordered systems in a uniform manner. It automatically adds precision based on the spurious traces it might generate.

*Outline.* Section (2) gives some preliminaries and defines parameterized systems. Section (3) describes the verification problem we target, and Section (4) introduces a generic verification to solve it. Section (5) introduces counted words and Section (6) uses them to instantiate the verification algorithm. Section (7) describes the experiments we performed and Section (8) concludes. Proofs and details of the examples are in the appendix.

## 2 Preliminaries

We use $\mathbb{N}$ for the set of natural numbers. For a natural number $n$, we use $\overline{n}$ to mean the set $\{1, \ldots, n\}$. We let $\Sigma^*$ be the set of finite words over $\Sigma$, $w \cdot w'$ be the concatenation of the words $w$ and $w'$, $\epsilon$ be the empty word, and $w \sqcup w'$ be the shuffle set $\{w'' | \ w'' = w_1 \cdot w'_1 \cdot w_2 \cdots w'_m \text{ s.t } w = w_1 \cdots w_n, w' = w'_1 \cdots w'_m\}$. Fix a word $w = \sigma_1 \cdots \sigma_n$. We write $|w|$ to mean the size $n$, $w_{[i,j]}$ to mean the word $\sigma_i \cdot \sigma_{i+1} \cdots \sigma_j$, $w_{[i]}$ to mean the letter $\sigma_i$, $hd(w)$ to mean the letter $\sigma_1$, and $tl(w)$ to mean the suffix $w_{[2,n]}$. We write $w^\bullet$ to mean the set $\{\sigma_1, \ldots, \sigma_n\}$ of letters appearing in $w$. A multiset $m$ is a mapping $\Sigma \to \mathbb{N}$. We write $m \subseteq m'$ to mean that $m$ is included in $m'$, i.e., that $\wedge_{\sigma \in \Sigma} m(\sigma) \leq m'(\sigma)$. We write $m \oplus m'$ to mean the union of $m$ and $m'$, i.e., $\wedge_{\sigma \in \Sigma}(m \oplus m')(\sigma) = m(\sigma) + m'(\sigma)$. If $m' \subseteq m$, then the multiset $m \ominus m'$ is defined and verifies, for each $\sigma$ in $\Sigma$, $(m \ominus m')(\sigma) = m(\sigma) - m'(\sigma)$. It is undefined otherwise. The Parikh image $w^\#$ of a word $w$ is the multiset that gives the number of occurrences in $w$ of each letter $\sigma$ in $\Sigma$. Given a set $\Sigma$ and a pre-order[1] $\preceq$ on $\Sigma$, a $(\Sigma, \preceq)$-antichain is an infinite sequence $\sigma_1, \sigma_2, \ldots$ of elements of $\Sigma$, with $\sigma_i \not\preceq \sigma_j$ if $i < j$. A pair $(\Sigma, \preceq)$ is a well quasi ordering if there are no $(\Sigma, \preceq)$-antichains.

## 3 Parameterized Systems with Global Conditions

Formally, a *parameterized system* is a pair $\mathcal{P} = (Q, T)$, where $Q$ is a finite set of *local states* and $T$ is a finite set of *transitions*. A transition is either *local* or *global*. A local transition is of the form $q \to q'$. It allows a process to change its local state from $q$ to $q'$ independently of the local states of the other processes. A global transition is of the form $q \to q' : \mathbb{Q}P$, where $\mathbb{Q} \in \{\exists_L, \exists_R, \exists_{LR}, \forall_L, \forall_R, \forall_{LR}\}$ and $P \subseteq Q$. Here, the process checks also the local states of the other processes when it takes the transition. For instance, the condition $\forall_L P$ means that "all processes to the left should be in local states which belong to the set $P$"; the condition $\forall_{LR}P$ means that "all other processes (whether to the left or to the right) should be in local states which belong to the set $P$". Given $Q$ and $T$, a parameterized system $\mathcal{P} = (Q,T)$ induces an infinite-state transition system $(C, \longrightarrow)$ where $C = Q^*$ is the set of *configurations* and $\longrightarrow$ is a transition relation on $C$. For configurations $c = c_1qc_2$, $c' = c_1q'c_2$, and a transition $t \in T$, we write $c \longrightarrow_t c'$ to mean that one of the following conditions is satisfied:

- $t$ is a local transition of the form $q \to q'$.
- $t$ is a global transition $q \to q' : \mathbb{Q}P$, and one of the following conditions is satisfied:
  - either $\mathbb{Q}P = \exists_L P$ and $c_1^\bullet \cap P \neq \emptyset$, or $\mathbb{Q}P = \exists_R P$ and $c_2^\bullet \cap P \neq \emptyset$, or $\mathbb{Q}P = \exists_{LR} P$ and $(c_1^\bullet \cup c_2^\bullet) \cap P \neq \emptyset$.
  - or $\mathbb{Q}P = \forall_L P$ and $c_1^\bullet \subseteq P$, or $\mathbb{Q}P = \forall_R P$ and $c_2^\bullet \subseteq P$, or $\mathbb{Q}P = \forall_{LR} P$ and $(c_1^\bullet \cup c_2^\bullet) \subseteq P$.

---

[1] i.e., a reflexive and transitive binary relation

We write $\longrightarrow$ to mean $\cup_{t \in T} \longrightarrow_t$ and use $\stackrel{*}{\longrightarrow}$ to denote the reflexive transitive closure of $\longrightarrow$. Given a parameterized system, we assume that, prior to starting the execution of the system, each process is in an (identical) *initial* state $p_{init}$. We use $Init$ to denote the set of *initial* configurations, i.e., configurations of the form $p_{init} \cdots p_{init}$ (all processes are in their initial states). Notice that the set $Init$ is infinite. It can be shown, using standard techniques (see e.g. [23]), that checking safety properties (expressed as regular languages) can be translated into instances of the reachability problem. The *reachability problem* for parameterized systems is defined as follows:

---
REACH-PAR
**Instance**
- A parameterized system $\mathcal{P} = (Q,T)$.
- A (possibly infinite) set $C_F$ of configurations.

**Question** $Init \stackrel{*}{\longrightarrow} C_F$ ?

---

## 4 A Generic Refinement Scheme

We introduce in this Section a generic scheme for solving the reachability problem of Section (3). The problem is undecidable in general. The scheme we introduce uses over-approximations to deduce unreachability. Each time the approximated analysis exhibits a sequence from the initial to the final configurations (i.e., a trace), we automatically follow the sequence in the original system. If it is possible we return it as a proof of reachability. Otherwise, the trace is a false positive and we automatically strengthen the approximation in order to prune the trace (Figure (1)).

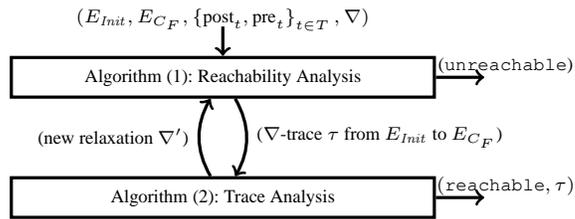

**Fig. 1.** A Generic scheme.

*Requirements on a symbolic representation.* A symbolic representation $\mathbb{S}$ permits to denote and to manipulate possibly infinite sets of configurations of a system $(Q, T)$. For an element $e$ in $\mathbb{S}$, we write $[\![e]\!]$ to mean the set of configurations denoted[2] by $e$. We write $[\![\{e_1, \ldots e_n\}]\!]$ to mean $\bigcup_{(i \text{ in } \overline{n})} [\![e_i]\!]$. In addition, we require that $\mathbb{S}$ verifies that:

1. We can effectively check whether $e'$ entails $e$ (write $e \sqsubseteq_\mathbb{S} e'$). We require that $\sqsubseteq_\mathbb{S}$ is reflexive and transitive, and that $e \sqsubseteq_\mathbb{S} e'$ implies $[\![e']\!] \subseteq [\![e]\!]$. We write $\{e_1, \ldots e_n\} \sqsubseteq_\mathbb{S} \{e'_1, \ldots e'_m\}$ to mean that for each $j : 1 \leq j \leq m$ there is an $e_i$ such that $e_i \sqsubseteq_\mathbb{S} e'_j$. In this case, observe that $[\![\{e'_1, \ldots e'_m\}]\!] \subseteq [\![\{e_1, \ldots e_n\}]\!]$.
2. We can effectively compute $e \sqcap_\mathbb{S} e' = \{e_1, \ldots e_n\}$ s.t. $[\![e]\!] \cap [\![e']\!] = [\![e \sqcap_\mathbb{S} e']\!]$. We simply let $\{e_1, \ldots e_n\} \sqcap_\mathbb{S} \{e'_1, \ldots e'_m\}$ be the set $\cup_{(i \text{ in } \overline{n}, j \text{ in } \overline{m})}(e_i \sqcap_\mathbb{S} e'_j)$.
3. $\square_\mathbb{S}$ is a set of effective relaxation operators. Each such operator $\nabla : \mathbb{S} \to \mathbb{S}$ verifies:
   (a) for each $e$ in $\mathbb{S}$, $\nabla(e) \sqsubseteq_\mathbb{S} e$, and
   (b) if $e \sqsubseteq_\mathbb{S} e'$, then $\nabla(e) \sqsubseteq_\mathbb{S} \nabla(e')$, and
   (c) there are no $(\mathbb{S}, \sqsubseteq_\mathbb{S})$-antichains, i.e., $(\mathbb{S}, \sqsubseteq_\mathbb{S})$ is a well quasi ordering.
4. $\Xi$ is an effective separation operator s.t. given finite $\{e_1, \ldots e_n\}$ and $\{e'_1, \ldots e'_m\}$ with $(\{e_1, \ldots e_n\} \sqcap_\mathbb{S} \{e'_1, \ldots e'_m\} = \emptyset)$, and given a relaxation $\nabla$ in $\square_\mathbb{S}$ with[3] $(\nabla(\{e_1, \ldots e_n\}) \sqcap_\mathbb{S} \{e'_1, \ldots e'_m\} \neq \emptyset)$, then $\Xi(\{e_1, \ldots e_n\}, \{e'_1, \ldots e'_m\}, \nabla)$ returns a stronger relaxation[4] $\nabla'$ in $\square_\mathbb{S}$ with $(\nabla'(\{e_1, \ldots e_n\}) \sqcap_\mathbb{S} \{e'_1, \ldots e'_m\} = \emptyset)$.
5. There are finite subsets $E_{Init}$ and $E_{C_F}$ such that $[\![E_{Init}]\!] = Init$ and $[\![E_{C_F}]\!] = C_F$.
6. For each $t$ in $T$ and $e$ in $\mathbb{S}$,
   (a) we can effectively compute sets $\text{post}_t(e)$ and $\text{pre}_t(e)$ such that $[\![\text{post}_t(e)]\!] = \{c' | c \longrightarrow_t c' \text{ s.t. } c \text{ in } [\![e]\!]\}$ and $[\![\text{pre}_t(e)]\!] = \{c' | c' \longrightarrow_t c \text{ s.t. } c \text{ in } [\![e]\!]\}$.
   (b) if $e \sqsubseteq_\mathbb{S} e'$, then $\text{post}_t(e) \sqsubseteq_\mathbb{S} \text{post}_t(e')$ and $\text{pre}_t(e) \sqsubseteq_\mathbb{S} \text{pre}_t(e')$

Requirements (1, 5, 6) are natural. Requirements (2, 3.b, 4) are needed by Algorithm (2), where (3.b) ensures the same trace is not encountered more than once. Requirement (3.a) is important for soundness of Algorithm (1), and requirement (3.c) guarantees its termination. We use $\mathbb{S}$ to implement the scheme of Figure (1).

*The reachability checking algorithm.* Algorithm (1) is a classical working list algorithm. It manipulates pairs $(e, \tau)$ of constraints and *traces*. A trace $\tau$ wrt. to a relaxation $\nabla$ (or a $\nabla$-trace for short) is a sequence $e_0 \cdot t_1 \cdot e_2 \cdots e_m$ of $\mathbb{S}$ elements $\{e_0, \ldots e_m\}$ and of transitions $\{t_1, \ldots t_m\}$ in $T$, such that $e_0$ in $E_{Init}$ and $e_{i+1} \in \nabla(\text{post}_{t_{i+1}}(e_i))$ for each $i : 0 \leq i < m$. Each manipulated pair is of the form $(e_m, e_0 \cdot t_1 \cdots e_m)$. The Algorithm maintains two sets $\mathcal{W}$ (working set) and $\mathcal{V}$ (visited set) such that $(\mathcal{W} \cup \mathcal{V})$ is minimal[5]. The set $\mathcal{W}$ collects pairs $(e, \tau)$ where $\text{post}_t(e)$ has still to be applied for each $t \in T$. The set $\mathcal{V}$ collects pairs $(e, \tau)$ where $\text{post}_t(e)$ has already been applied for each transition $t$ in $T$. Initially, no element of $\mathbb{S}$ is visited, and all members of $E_{Init}$ (assumed minimal) are added to the working set (line (1)). If there is a pair $(e_c, \tau)$ in the working set, it is first removed from $\mathcal{W}$ (line (3)). If its denotation intersects $C_F$, then we found a trace in the over-approximated system from the initial to the final configurations. In this case, the Algorithm returns $\tau$ as a proof of reachability (lines (4,5)). Otherwise, the pair is added to the visited set (line (6)) and $\text{post}_t(e_c)$ is computed for each $t$ in $T$. Each element in $\text{post}_t(e_c)$ is relaxed (line (8)) before being added to the set $New_t$. This relaxation is at the source of imprecision and will guarantee termination. Elements of $New_t$ are pruned away if they do not add new configurations. Otherwise, they are used to remove redundant elements of $\mathcal{V} \cup \mathcal{W}$ before being added to $\mathcal{W}$ together with their updated traces (lines (11,12)).

---
[2] to simplify, we assume $[\![e]\!] \neq \emptyset$ for each $e$ in $\mathbb{S}$
[3] we write $\nabla(\{e_1, \ldots e_n\})$ to mean the set $\{\nabla(e_1), \ldots \nabla(e_m)\}$, and let $\nabla(\{\}) = \{\}$.
[4] $\nabla'$ is stronger than $\nabla$ if $\nabla(e) \sqsubseteq_\mathbb{S} \nabla'(e)$ for each $e$ in $\mathbb{S}$
[5] $\{e_1, \ldots e_n\}$ is minimal if $e_i \not\sqsubseteq_\mathbb{S} e_j$ if $i \neq j$.

**Algorithm 1:** The reachability checker

**input** : $E_{Init}$ and $E_{C_F}$, operators $\text{post}_t(.)$ and $\text{pre}_t(.)$ for each $t \in T$, and a relaxation $\nabla$
**output**: a *trace* $(e_0 \cdot t_1 \cdot e_2 \cdot t_2 \cdots e_m)$ with $\{e_m\} \sqcap_\mathbb{S} E_{C_F} \neq \emptyset$, or unreachable

1. $\mathcal{W} := \{(e, e) | \ e \text{ in } E_{Init}\}, \mathcal{V} := \{\}$;
2. **while** $(\mathcal{W} \neq \{\})$ **do**
3.   Pick and remove a pair $(e_c, \tau)$ from $\mathcal{W}$;
4.   **if** $(\{e_c\} \sqcap_\mathbb{S} E_{C_F} \neq \emptyset)$ **then**
5.     **return** $\tau$;
6.   $\mathcal{V} := \{(e_c, \tau)\} \cup \mathcal{V}$;
7.   **foreach** $t \in T$ **do**
8.     $New_t := \{\nabla(e) | \ e \in \text{post}_t(e_c)\}$;
9.     **foreach** $e \in New_t$ **do**
10.       **if** $\forall (e_{old}, \tau_{old}) \in \mathcal{W} \cup \mathcal{V}. \ e_{old} \not\sqsubseteq_\mathbb{S} e$ **then**
11.         $\mathcal{V} := \{(e_{old}, \tau_{old}) | \ (e_{old}, \tau_{old}) \in \mathcal{V} \wedge e \not\sqsubseteq_\mathbb{S} e_{old}\}$;
12.         $\mathcal{W} := \{(e, \tau \cdot t \cdot e)\} \cup \{(e_{old}, \tau_{old}) | \ (e_{old}, \tau_{old}) \in \mathcal{W} \wedge e \not\sqsubseteq_\mathbb{S} e_{old}\}$
13. **return** unreachable;

**Lemma 1 (reachability).** *Algorithm (1) always terminates. If it returns* unreachable, *then* $(Init \xrightarrow{*} C_F)$ *does not hold for the parameterized system* $\mathcal{P} = (Q, T)$. *Otherwise, it returns a trace* $e_0 \cdot t_1 \cdot e_1 \cdots e_m$ *with* $e_0 \in E_{Init}$, $\{e_m\} \sqcap_\mathbb{S} E_{C_F} \neq \emptyset$, *and* $e_{i+1} \in \nabla(\text{post}_{t_{i+1}}(e_i))$ *for each* $i : 0 \leq i < m$.

*The trace analyzer.* Algorithm (2) simulates a trace backwards[6] in order to check its possibility in the original system. Only the approximation resulting from the applications of the $\nabla$ relaxation can result in the analyzer failing to follow the supplied $\nabla$-trace. If this happens (lines (4) or (10)), the analyzer relies on the separation operator (lines (5) and (11)) to supply a stronger relaxation operator that will prune the trace in future analysis. Otherwise, the analyzer will manage to reach the initial configurations (line (8)). In this case, it returns the trace as a proof of reachability (line (9)).

**Algorithm 2:** The trace analyzer

**input** : $\nabla$-trace $(e_0 \cdot t_1 \cdot e_2 \cdots e_m)$ with $\{e_m\} \sqcap_\mathbb{S} E_{C_F} \neq \emptyset$
**output**: reachable with input trace, or a relaxation operator $\nabla'$

1. $Current := \{e_m\} \sqcap_\mathbb{S} E_{C_F}$;
2. **foreach** $i = m - 1$ *to* $0$ **do**
3.   $Predecessor := (\bigcup_{e \in Current} \text{pre}_{t_{i+1}}(e)) \sqcap_\mathbb{S} \{e_i\}$;
4.   **if** $(Predecessor = \emptyset)$ **then**
5.     **return** $\Xi(\text{post}_{t_i}(e_i), Current, \nabla)$;
6.   **else** $Current := Predecessor$
7. $Inter := Current \sqcap_\mathbb{S} E_{Init}$;
8. **if** $Inter \neq \emptyset$ **then**
9.   **return** reachable, $\tau$
10. **else**
11.   **return** $\Xi(\{e_0\}, E_{Init}, \nabla)$

**Lemma 2 (Refinement).** *Given a $\nabla$-trace* $\tau = e_0 \cdot t_1 \cdots e_m$ *with* $\{e_m\} \sqcap_\mathbb{S} E_{C_F} \neq \emptyset$, *Algorithm (2) terminates. If it returns* (reach, $\tau$), *then there are* $c_0, \ldots c_m$ *in* $C$, *with* $c_0 \in Init$, $c_m \in C_F$ *and s.t.* $c_i \longrightarrow_{t_{i+1}} c_{i+1}$ *for* $i : 0 \leq i < m$. *Otherwise, it returns a stronger relaxation* $\nabla'$ *such that no relaxation* $\nabla''$ *that is stronger than* $\nabla'$ *can have a* $\nabla''$*-trace* $e'_0 \cdot t_1 \cdots e'_m$ *with* $\{e'_m\} \sqcap_\mathbb{S} E_{C_F} \neq \emptyset$ *and where* $e_i \sqsubseteq_\mathbb{S} e'_i$ *for* $i : 0 \leq i \leq m$.

By combining Lemmata (1) and (2) and the requirement that $\Xi$ returns a stronger relaxation operator, we get correctness of the Algorithm depicted by Figure (1).

**Theorem 1.** *Each iteration of the Algorithm depicted in Figure (1) terminates. If it returns* unreachable, *then* $Init \xrightarrow{*} C_F$ *does not hold. If it returns* (reachable, $\tau$), *then* $Init \xrightarrow{*} C_F$ *via the transitions in* $\tau$. *In addition, no trace is generated twice.*

---
[6] here in a forward analysis. For a backward analysis, switch $E_{Init}, E_{C_F}, \text{post}_{(.)}(.), \text{pre}_{(.)}(.)$ respectively with $E_{C_F}, E_{Init}, \text{pre}_{(.)}(.), \text{post}_{(.)}(.)$ in both Algorithms (1) and (2).

## 5 Counted Words

*Counters.* We fix a finite set of variables $V_Q$. We define in the following the set of counters $\mathbb{C}$ over $Q$. The set $V_Q$ is in a one to one correspondence with $Q$. Each variable $v$ is associated to a state $q$ in $Q$. We write $v_q$ to make the association clear. Intuitively, $v_q$ is used to count the number of occurrences of the associated letter $q$ in a word in $Q^*$. A counter basically captures multisets over $Q$ by separately imposing a constraint on each state in $Q$. Indeed, we define a counter $cr$ to be a conjunction $[\wedge_{q \in Q}(v_q \sim k)]$ where $\sim$ is in $\{=, \geq\}$, each $v_q$ is a variable ranging over $\mathbb{N}$ and each $k$ is a constant in $\mathbb{N}$. For a state $q$ in $Q$, we write $cr(q)$ to mean the strongest predicate of the form $(v_q \sim k)$ implied by the counter $cr$. We write $cr_q$ to mean the counter $[\wedge_{q_i \in Q}(v_{q_i} = b_{q_i})]$ with $b_{q_i}$ equal to 1 for $q_i = q$ and to 0 otherwise. A substitution is a set $\{v_1 \leftarrow u_1, \ldots\}$ of pairs[7] where $v_1, \ldots$ are variables, and $u_1, \ldots$ are either all variables or all natural numbers. Given a counter $cr$ and a substitution $S$, we write $cr[S]$ to mean the formula obtained by replacing, for each pair $v_i \leftarrow u_i$, each occurrence of $v_i$ in $cr$ by $u_i$. We sometimes regard a multiset $m$ as the substitution $\{v_q \leftarrow m(q)|\ q \text{ in } Q\}$. For a counter $cr$ and a multiset $m$, the formula $cr[m]$ takes a Boolean value. In the case where it evaluates to true (resp. false), we say that $m$ satisfies (resp. doesn't satisfy) the counter $cr$. Given a word $w$ in $Q^*$ and a counter $cr$, we abuse notation and write $cr[w]$ to mean that $(w^\#)$ satisfies $cr$. For a counter $cr$, we write $[\![cr]\!]$ to mean the set $\{w|\ cr[w] \text{ and } w \in Q^*\}$. We define the precision of a counter $cr$, written $\kappa(cr)$, to be the multiset that associates to each state $q$ in $Q$ the value 0 if $cr(q) = (v_q \geq k)$, and $k+1$ if $cr(q) = (v_q = k)$. Observe that if $\kappa(cr)(q) \neq 0$ for all $q \in Q$, then $cr$ accepts a single multiset, while if $\kappa(cr)(q) = 0$ for all $q \in Q$, then $cr$ accepts an upward closed set of multisets (wrt. $\subseteq$). For a natural number $k$, we write $\mathbb{C}_k$ to mean the set $\{cr|\ \kappa(cr)(q) \leq k \text{ for each } q \in Q\}$.

*Example 1.* Assume a counter $cr = [v_a = 0 \wedge v_b = 2 \wedge v_c \geq 1]$. The following holds: $\kappa(cr)(a) = 1, \kappa(cr)(b) = 3$, and $\kappa(cr)(c) = 0$. In addition, $cr$ is in $\mathbb{C}_3$.

We use $\bot_\mathbb{C}$ to mean $[false]$ ($[\![\bot_\mathbb{C}]\!] = \emptyset$) and $\top_\mathbb{C}$ to mean $[\wedge_{q \in Q}(v_q \geq 0)]$ ($[\![\top_\mathbb{C}]\!] = Q^*$). We assume two counters $cr$ and $cr'$ and define a number of operations on them.

- The meet $cr \sqcap_\mathbb{C} cr'$ is the conjunction $cr \wedge cr'$ of the two counters. The denotation of the meet $[\![cr \sqcap_\mathbb{C} cr']\!]$ is the intersection of the denotations $[\![cr]\!] \cap [\![cr']\!]$.
- The counter $cr'$ is said to entail the counter $cr$ (we write $cr \sqsubseteq_\mathbb{C} cr'$) if $cr'$ implies $cr$, i.e, $cr' \Rightarrow cr$. Observe that $cr \sqsubseteq_\mathbb{C} cr'$ means $[\![cr']\!] \subseteq [\![cr]\!]$.

$$\exists \left(v'_{q_1}, v''_{q_1} \ldots v'_{q_n}, v''_{q_n}\right) \cdot \left(\bigwedge_{q_i \in Q} v_{q_i} = v'_{q_i} + v''_{q_i}\right) \wedge cr[S'] \wedge cr'[S''] \quad (3)$$

$$\exists \left(v'_{q_1}, v''_{q_1} \ldots v'_{q_n}, v''_{q_n}\right) \cdot \left(\bigwedge_{q_i \in Q} v_{q_i} = v'_{q_i} - v''_{q_i} \wedge v_{q_i} \geq 0\right) \wedge cr[S'] \wedge cr'[S''] \quad (4)$$

- The sum $cr \oplus_\mathbb{C} cr'$ is the conjunction obtained in (3), with $S' = \{v_q \leftarrow v'_q|\ q \in Q\}$ and $S'' = \{v_q \leftarrow v''_q|\ q \in Q\}$. Intuitively, $[\![cr \oplus_\mathbb{C} cr']\!]$ coincides with the shuffle set $\{w \sqcup w'|\ w \in [\![cr]\!] \text{ and } w' \in [\![cr']\!]\}$. In fact, $(cr_1 \oplus_\mathbb{C} cr_2)[m]$ iff there are multisets $m_1, m_2$ with $m = m_1 \oplus m_2$ and s.t. $cr_1[m_1]$ and $cr_2[m_2]$.

---
[7] we assume variables appearing to the left are distinct, i.e. $v_i \neq v_j$ if $i \neq j$.

– In a similar manner, the difference $cr \ominus_{\mathbb{C}} cr'$ is the conjunction obtained in (4), where $S', S''$ are defined as above. Intuitively, $cr \ominus_{\mathbb{C}} cr'$ denotes words that can be shuffled with a word from $[\![cr']\!]$ to obtain a word in $[\![cr]\!]$. That is, $[\![cr \ominus_{\mathbb{C}} cr']\!]$ is $\{w|\; w \sqcup w' \in [\![cr]\!]$ and $w' \in [\![cr']\!]\}$. In other words, $(cr_1 \ominus_{\mathbb{C}} cr_2)[m]$ iff there are multisets $m_1$ and $m_2$ with $m = m_1 \ominus m_2$ defined and s.t. $cr_1[m_1]$ and $cr_2[m_2]$.

**Lemma 3.** *For any $k \in \mathbb{N}$, $(\mathbb{C}_k, \sqsubseteq_{\mathbb{C}})$ is a well quasi ordering. In fact, from every infinite sequence $(cr_1, cr_2, \ldots)$, we can extract an infinite subsequence $(cr_{i_1} \sqsubseteq_{\mathbb{C}} cr_{i_2} \sqsubseteq_{\mathbb{C}} \ldots)$.*

*Counted words.* A counted word is any member $\varphi$ in $(\mathbb{C} \times Q \times \mathbb{C})^*$. If $(l, q, r) \in (\mathbb{C} \times Q \times \mathbb{C})$, we write $(l, q, r)_{st}$ to mean $q$. Assume $\varphi = (l_1, q_1, r_1) \cdots (l_n, q_n, r_n)$ is a counted word. The *base* of $\varphi$ (written $\underline{\varphi}$) is the word $q_1 \cdots q_n$ in $Q^*$. We write $lc(\varphi)$ (resp. $rc(\varphi)$) to mean the counter $\top_{\mathbb{C}}$ if $\overline{\varphi} = \epsilon$, and $l_1$ (resp. $r_n$) otherwise. We refer to $l_1, \ldots l_n$ (resp. $r_1, \ldots r_n$) as the left (resp. right) counters of $\varphi$. The counted word $\varphi$ is *well formed* if $l_i[(\underline{\varphi})_{[1,i-1]}]$ and $r_i[(\underline{\varphi})_{[i+1,n]}]$ for each $i : 1 \le i \le n$. We assume $\epsilon$ is *well formed*. Example (2) depicts a counted word. Well formedness imposes predicates in the counters are of a certain form. This is captured by the following lemma.

**Lemma 4 (Well formedness).** *Assume a counted word $\varphi = (l_1, q_1, r_1) \cdots (l_n, q_n, r_n)$. For each $i : 1 \le i \le n$, let $p_i^l = ((\underline{\varphi})_{[1,i-1]})^{\#}$ and $p_i^r = ((\underline{\varphi})_{[i+1,n]})^{\#}$. Then:*

- *Each $l_i(q)$ equals $(v_q = (p_i^l(q)))$ or $(v_q \ge k)$ for some $k$ in $\{0, \ldots (p_i^l(q))\}$.*
- *Each $r_i(q)$ equals $(v_q = (p_i^r(q)))$ or $(v_q \ge k)$ for some $k$ in $\{0, \ldots (p_i^r(q))\}$.*

*Denotation.* Given a word $w = q_1 \cdots q_n$ and an increasing injection $h : \overline{n} \to \overline{m}$, we write $w \models^h \varphi$ to mean that all following three conditions hold for each $i : 1 \le i \le n$
i) $\underline{\varphi}[i] = w[h(i)]$, and ii) $l_i(w_{[1,h(i)-1]})$, and iii) $r_i(w_{[h(i)+1,n]})$. Intuitively, there is an injection $h$ that ensures $\underline{\varphi}$ is subword of $w$, and s.t. words to the left and right of each image of $h$ respectively respect corresponding left and right counters in $\varphi$. We write $w \models \varphi$ if $w \models^h \varphi$ for some injection $h$, and $[\![\varphi]\!]$ to mean $\{w|\; w \models \varphi\}$. We let $[\![\epsilon]\!] = Q^*$. Observe that every well formed word has a non-empty denotation since $\underline{\varphi} \models \varphi$. We use $\mathbb{CW}$ to mean the set of well formed counted words.

*Example 2.* $\varphi = \left( \begin{bmatrix} v_a = 0 \\ \wedge v_b \ge 0 \end{bmatrix}, a, \begin{bmatrix} v_a \ge 0 \\ \wedge v_b \ge 0 \end{bmatrix} \right) \left( \begin{bmatrix} v_a = 1 \\ \wedge v_b = 0 \end{bmatrix}, a, \begin{bmatrix} v_a = 0 \\ \wedge v_b \ge 0 \end{bmatrix} \right)$ and $[\![\varphi]\!] = aab^*$.

*Entailment.* For $\varphi = (l_1, q_1, r_1) \cdots (l_n, q_n, r_n)$ and $\varphi' = (l'_1, q'_1, r'_1) \cdots (l'_m, q'_m, r'_m)$, we say that $\varphi$ is entailed by $\varphi'$ if $\varphi \sqsubseteq_{\mathbb{CW}}^h \varphi'$ for some injection $h : \overline{n} \to \overline{m}$; where $\varphi \sqsubseteq_{\mathbb{CW}}^h \varphi'$ requires for each $i : 1 \le i \le n$, that the following three conditions hold: $\underline{\varphi}_{[i]} = \underline{\varphi'}_{[h(i)]}$, $l_i \sqsubseteq_{\mathbb{C}} l'_{h(i)}$, and $r_i \sqsubseteq_{\mathbb{C}} r'_{h(i)}$. We write $\varphi \sqsubseteq_{\mathbb{CW}} \varphi'$ to mean that $\varphi$ is entailed by $\varphi'$. Observe that $([v_a \ge 0], a, [v_a = 0]) \not\sqsubseteq_{\mathbb{CW}} ([v_a = 0], a, [v_a \ge 0])$, but $[\![([v_a \ge 0], a, [v_a = 0])]\!] = [\![([v_a = 0], a, [v_a \ge 0])]\!]$.

**Lemma 5 (Entailment).** *The relation $\sqsubseteq_{\mathbb{CW}}$ is both reflexive and transitive. In addition, $\varphi \sqsubseteq_{\mathbb{CW}} \varphi'$ implies $[\![\varphi']\!] \subseteq [\![\varphi]\!]$.*

*Bounded precision.* We define the precision of a well formed word $\varphi$ as a multiset $\kappa(\varphi)$. It associates to each $q$ the natural number $max\ \{\kappa(cr)(q)|\ cr$ is a counter in $\varphi\}$. In Example (2) for instance, $\kappa(\varphi)(a) = 2$ and $\kappa(\varphi)(b) = 1$. We say that a counted word $\varphi$ has a $k$-bounded precision if all its counters are in $\mathbb{C}_k$. For example, counted words with a $0$-bounded precision only have inequalities in their counters (they denote upward closed sets with respect to the subword ordering). We write $\mathbb{CW}_k$ to mean the set of well formed counted words that have a $k$-bounded precision.

**Theorem 2 (WQO).** *For any fixed $k \in \mathbb{N}$, $(\mathbb{CW}_k, \sqsubseteq_{\mathbb{CW}})$ is a well quasi ordering.*

*Strengthening of well formed words.* Counters in a counted word are not independent. Consider for instance $\varphi = (l_1, a, r_1)(l_2, a, r_2)$ in Example (2). We can change $l_1(b)$ to $(v_b = 0)$ without affecting the denotation of $\varphi$. The reason is that any prefix accepted by $l_1$ will have to be allowed by $l_2$. It is therefore vacuous for $l_1$ to accept words containing $b$, and more generally to accept more than $l_2 \ominus_{\mathbb{C}} cr_a$. Also, observe that $l_2$ and $r_2$ imply we can change $r_1(a)$ from $(v_a \geq 0)$ to $(v_a = 1)$. We strengthen the counters of a well formed word by applying in any order rules in Figure (2) until a fixpoint is reached.

$$\frac{\varphi_p \cdot (l,q,r) \cdot (l',q',r') \cdot \varphi_s}{\varphi_p \cdot (l,q,r \sqcap r'') \cdot (l',q',r') \cdot \varphi_s} \text{ right} \qquad \frac{\varphi_p \cdot (l,q,r) \cdot (l',q',r') \cdot \varphi_s}{\varphi_p \cdot (l,q,r) \cdot (l',q',r' \sqcap (r \ominus_{\mathbb{C}} cr_{q'})) \cdot \varphi_s} \text{ right}'$$

$$\frac{\varphi_p \cdot (l,q,r) \cdot (l',q',r') \cdot \varphi_s}{\varphi_p \cdot (l \sqcap (l' \ominus_{\mathbb{C}} cr_q), q, r) \cdot (l',q',r') \cdot \varphi_s} \text{ left} \qquad \frac{\varphi_p \cdot (l,q,r) \cdot (l',q',r') \cdot \varphi_s}{\varphi_p \cdot (l,q,r) \cdot ((l' \sqcap l''), q', r') \cdot \varphi_s} \text{ left}'$$

**Fig. 2.** Strengthening rules for counted words. The counter $r''$ in rule right equals $r' \oplus_{\mathbb{C}} cr_{q'} \oplus_{\mathbb{C}} (l' \ominus_{\mathbb{C}} (l \oplus_{\mathbb{C}} cr_q))$, and the counter $l''$ in rule left' equals $l \oplus_{\mathbb{C}} cr_q \oplus_{\mathbb{C}} (r \ominus_{\mathbb{C}} (r' \oplus_{\mathbb{C}} cr_{q'}))$.

**Lemma 6 (Strengthening).** *Given a well formed word $\varphi$, the strengthening procedure terminates and yields a unique well formed word $\varphi'$ s.t. $[\![\varphi]\!] = [\![\varphi']\!]$ and $\varphi \sqsubseteq_{\mathbb{CW}} \varphi'$.*

Let $\mathbb{SCW}$ (resp. $\mathbb{SCW}_k$) be the set of strengthened words in $\mathbb{CW}$ (resp. in $\mathbb{CW}_k$). We will use $\mathbb{SCW}$ as a symbolic representation for the generic scheme of Section (4).

## 6 Instantiation of the Refinement Algorithm

We instantiate the scheme of Section (4) using the set $\mathbb{SCW}$ as a symbolic representation. For this, we define a family of relaxation operators, show how to compute successors and predecessors on $\mathbb{SCW}$, and describe both meet and separation operators.

*Relaxation.* We use the notion of relaxing a counted word $\varphi$ wrt. a resolution (in this context, a multiset) $\rho$. First, given a counter $cr$, relaxation of $cr$ wrt. to $\rho$, written $\nabla_\rho(cr)$, is the counter $[\wedge_{q \text{ in } Q}(v_q \sim k)]$ s.t. $(v_q \sim k)$ is equal to $(v_q = k)$ if $cr(q)$ was $(v_q = k)$ with $k < \rho(q)$, and equal to $v_q \geq k$ otherwise. In other words, relaxation does not keep track of equalities larger than what is allowed by the resolution. Relaxation of a counted word $\varphi$ wrt. a resolution $\rho$ is simply the word $\nabla_\rho(\varphi)$ obtained by strengthening the word resulting from relaxation of all counters in $\varphi$ wrt. $\rho$. We let $\Box_{\mathbb{SCW}}$ be the set $\{\nabla_\rho|\ \rho$ is a multiset over $Q\}$.

**Lemma 7 (Relaxation).** *Given $\varphi$ in $\mathbb{CW}$ and resolutions $\rho \subseteq \rho'$, we have: $\nabla_\rho(\varphi) \sqsubseteq_{\mathbb{CW}} \nabla_{\rho'}(\varphi) \sqsubseteq_{\mathbb{CW}} \varphi$ and $\nabla_\rho(\varphi)$ is in $\mathbb{SCW}_{max(0,2k-1)}$ with $k = max\{\rho(q)|\ q\ in\ Q\}$.*

*Post and Pre operators.* First, we define an operator $q \otimes \varphi$ that takes a strengthened well formed word $\varphi$ and a state $q$ and returns all tuples $(\varphi_1, (l,q,r), \varphi_2)$ s.t. either $\varphi = \varphi_1 \cdot (l,q,r) \cdot \varphi_2$, or $\varphi = \varphi_1 \cdot \varphi_2$ with[8] $q \in cxt(rc(\varphi_1)) \cap cxt(lc(\varphi_2))$, $l = lc(\varphi_2)$, and $r = rc(\varphi_1)$. If $(\varphi_1, (l,q,r), \varphi_2)$ is in $q \otimes \varphi$, then $\varphi \sqsubseteq_{\mathbb{CW}} (\varphi_1 \cdot (l,q,r) \cdot \varphi_2)$. Intuitively, if it is possible to place the state $q$ in some position in $\varphi$, there will be a tuple $(\varphi_1, (l,q,r), \varphi_2)$ in $q \otimes \varphi$ to capture that. In addition, for $P \subseteq Q$, we write $0_P$ to mean the counter $[\wedge_{q \in P}(v_q = 0) \wedge \wedge_{q \notin P}(v_q \geq 0)]$. We describe how to compute $post_t(\varphi)$ and $pre_t(\varphi)$ for a transition $t \in T$ and a word $\varphi$ in $\mathbb{SCW}$. For each local $(q \to q')$, or global $(q \to q' : \mathbb{Q}P)$ transition $t$, the set $post_t(\varphi)$ is the smallest set containing strengthenings of all words $\varphi_1' \cdot (l', q', r') \cdot \varphi_2'$ that satisfy the following. There is a tuple $(\varphi_1, (l,q,r), \varphi_2)$ in $q \otimes \varphi$ s.t. $\varphi_1 = (l_1, q_1, r_1) \cdots (l_n, q_n, r_n)$ and $\varphi_2 = (l_{n+1}, q_{n+1}, r_{n+1}) \cdots (l_m, q_m, r_m)$, and:

1. Either $t$ is a local transition $q \to q'$, and $\varphi_1' = (l_1, q_1, r_1') \cdots (l_n, q_n, r_n')$ and $\varphi_2' = (l_{n+1}', q_{n+1}, r_{n+1}) \cdots (l_m', q_m, r_m)$ with $r_i' = r_i \oplus_{\mathbb{C}} cr_{q'} \ominus_{\mathbb{C}} cr_q$ for each $i : 1 \leq i \leq n$, and $l_i' = l_i \oplus_{\mathbb{C}} cr_{q'} \ominus_{\mathbb{C}} cr_q$ for each $i : n+1 \leq i \leq m$. Also, $l = l'$ and $r = r'$. Intuitively, we update the right counters of $\varphi_1$ and the left counters of $\varphi_2$ by requiring from all accepted multisets to have one less $q$ and one additional $q'$, or

2. $t$ equals $q \to q' : \forall_L P$, s.t. $(q_1 \cdots q_n)^\bullet \subseteq P$ and $\varphi_1' = (l_1', q_1, r_1') \cdots (l_n', q_n, r_n')$, $l' = l \sqcap_{\mathbb{S}} 0_{Q \setminus P}$, $r' = r$, and $\varphi_2' = (l_{n+1}', q_{n+1}, r_{n+1}) \cdots (l_m', q_m, r_m)$ with $l_i' = l_i \sqcap_{\mathbb{S}} 0_{Q \setminus P}$ and $r_i' = r_i \oplus_{\mathbb{C}} cr_{q'} \ominus_{\mathbb{C}} cr_q$ for each $i : 1 \leq i \leq n$, and $l_i' = l_i \oplus_{\mathbb{C}} cr_{q'} \ominus_{\mathbb{C}} cr_q$ for each $i : n+1 \leq i \leq m$. Intuitively, we check first that there is at least a (possibly empty) prefix in $P$. If it is the case, we require that all accepted multisets to the left of $q$ only contain states in $P$. In addition, we update the right counters of $\varphi_1$ and the left counters of $\varphi_2$ like in the previous case, or

3. $t$ is of the form $q \to q' : \exists_L P$, there is a tuple $(\varphi_1'', (l'', p, r''), \varphi_2'')$ in $p \otimes \varphi_1$ with $\varphi_1'' \cdot (l'', p, r'') \cdot \varphi_2'' = (l_1'', q_1'', r_1'') \cdots (l_{m''}'', q_{m''}, r_{m''}'')$, and $\varphi_1' = (l_1'', q_1'', r_1'' \oplus_{\mathbb{C}} cr_{q'} \ominus_{\mathbb{C}} cr_q) \cdots (l_{m''}'', q_{m''}, r_{m''}'' \oplus_{\mathbb{C}} cr_{q'} \ominus_{\mathbb{C}} cr_q)$ and $\varphi_2' = (l_{n+1}' \oplus_{\mathbb{C}} cr_{q'} \ominus_{\mathbb{C}} cr_q, q_{n+1}, r_{n+1}) \cdots (l_m' \oplus_{\mathbb{C}} cr_{q'} \ominus_{\mathbb{C}} cr_q, q_m, r_m)$. Also, $l = l'$ and $r = r'$. Intuitively, we make sure there is a witness $p$ in $P$ to the left of $q$. Then, we update the counters like for the first case.

The cases $q \to q' : \mathbb{Q}P$ where $\mathbb{Q}P$ is of the form $\forall_R P$ or $\forall_{LR} P$ are similar to case (2), and those where $\mathbb{Q}P$ is of the form $\exists_R P$ or $\exists_{LR} P$ are similar to case (3). Also, we let $pre_t(\varphi)$ be $post_{(q' \to q)}(\varphi)$ if $t = (q \to q')$ and $post_{(q' \to q:\mathbb{Q}P)}(\varphi)$ if $t = (q \to q' : \mathbb{Q}P)$.

**Lemma 8 (Post and Pre).** *Given a strengthened word $\varphi$ and a transition $t$ we can compute a set of words $post_t(\varphi)$ (resp. $pre_t(\varphi)$) such that $post_t(\varphi) \sqsubseteq_{\mathbb{CW}} post_t(\varphi')$ (resp. $pre_t(\varphi) \sqsubseteq_{\mathbb{CW}} pre_t(\varphi')$) if $\varphi \sqsubseteq_{\mathbb{CW}} \varphi'$. In addition, $[\![post_t(\varphi)]\!]$ and $[\![pre_t(\varphi)]\!]$ respectively equal $\{c'|\ c \longrightarrow_t c'\ with\ c\ in\ [\![\varphi]\!]\}$, and $\{c'|\ c' \longrightarrow_t c\ with\ c\ in\ [\![\varphi]\!]\}$.*

*Meet of counted words.* Given $\varphi, \varphi'$ in $\mathbb{SCW}$, we strengthen the result of Procedure (zip) and obtain a set $\varphi \sqcap_{\mathbb{CW}} \varphi'$ of counted words that entail both $\varphi$ and $\varphi'$ and whose denotation coincides with $[\![\varphi]\!] \cap [\![\varphi']\!]$. The procedure builds a constrained shuffle of $\varphi$ and $\varphi'$.

---

[8] we write $cxt(cr)$, for a counter $cr$, to mean the set $\{q|\ \kappa(cr)(q) = 0\}$.

It is recursive and takes as arguments five counted words $z, p, s, p', s'$, with $\varphi = (p \cdot s)$ and $\varphi' = (p' \cdot s')$. We write $(z, (p:s), (p':s'))$ for clarity. Intuitively, each call tries to complete the first argument $z$ in order to obtain a word that entails both $(p \cdot s)$ and $(p' \cdot s')$. The pro-

**Procedure** zip($z$, ($p$:$s$), ($p'$:$s'$))

```
1  collect := ∅;
2  if (s ≠ ε) then
3    if (hd(s)_st ∈ (cxt(rc(p')) ∩ cxt(lc(s')))) then
4      collect ∪ := zip((z · hd(s)), ((p · hd(s)) : tl(s)), (p' : s'))
5  if (s ≠ ε ∧ s' ≠ ε) then
6    if (lc(hd(s)) ⊓_S lc(hd(s')) ≠ ⊥_C) ∧ (hd(s)_st = hd(s')_st) ∧
       (rc(hd(s)) ⊓_S rc(hd(s')) ≠ ⊥_C) then
7      e := (lc(hd(s)) ⊓_S lc(hd(s')), hd(s)_st, rc(hd(s)) ⊓_S rc(hd(s')));
8      collect ∪ := zip((z · e) : ((p · hd(s)), tl(s)) :
            ((p' · hd(s')), tl(s')))
9  if (s' ≠ ε) then
10   if (hd(s')_st ∈ (cxt(rc(p)) ∩ cxt(lc(s)))) then
11     collect ∪ := zip((z · hd(s')) : (p, s) : ((p' · hd(s')), tl(s')))
12 if (s = ε ∧ s' = ε) then
13   collect := {z}
14 return collect;
```

cedure starts with $(\epsilon, (\epsilon : \varphi), (\epsilon : \varphi'))$ and collects all such words $z$. At each call, it considers contributions to $z$ from $hd(s)$ (lines (2-4)), $hd(s')$ (lines (9-11)), or both $hd(s)$ and $hd(s')$ (lines (5-8)). Lines (2-4) capture the situation where a state in $z$ is mapped to $hd(s)$ and tolerated by $\varphi'$ (test at line (3)). Lines (5-8) correspond to a state in $z$ simultaneously mapped to $hd(s)$ and $hd(s')$. The words $s$ and $s'$ contain states that are still not treated. Termination is obtained with the ranking function $|s| + |s'|$. The following lemma establishes correctness of Procedure (zip).

**Lemma 9 (intersection).** *Given $\varphi, \varphi'$ in $\mathbb{SCW}$,* zip($\epsilon : (\epsilon, \varphi) : (\epsilon, \varphi')$) *returns a set* $\{\varphi_1, \ldots \varphi_n\}$ *s.t.* $(\varphi \sqsubseteq_{\mathbb{CW}} \varphi_i)$, $(\varphi' \sqsubseteq_{\mathbb{CW}} \varphi_i)$ *for each $i \in \overline{n}$, and $\cup_{i \, in \, \overline{n}} [\![\varphi_i]\!] = [\![\varphi]\!] \cap [\![\varphi']\!]$.*

*Separation operator.* Assume strengthened words $\varphi, \varphi'$ and $\rho$ s.t. $(\varphi \sqcap_{\mathbb{CW}} \varphi' = \emptyset)$ but $(\nabla_\rho(\varphi) \sqcap_{\mathbb{CW}} \varphi' \neq \emptyset)$. We describe the operator $\Xi(\{\varphi\}, \{\varphi'\}, \rho)$.

**Procedure** augzip($z$, ($p$:$s$), ($\widehat{p}$:$\widehat{s}$), ($p'$:$s'$))

```
1  collect, avoid := ∅, true;
2  if (s ≠ ε) then
3    if (hd(s)_st ∈ (cxt(rc(p')) ∩ cxt(lc(s')))) then
4      c, v := augzip((z · hd(s)) : ((p · hd(s)), tl(s)) :
              ((p̂ · hd(ŝ)), tl(ŝ)) : (p', s'));
5      collect ∪ := c; avoid ∧ := v;
6  if (s ≠ ε ∧ s' ≠ ε) then
7    if (lc(hd(s)) ⊓_S lc(hd(s')) ≠ ⊥_C) ∧ (hd(s)_st = hd(s')_st) ∧
       (rc(hd(s)) ⊓_S rc(hd(s')) ≠ ⊥_C) then
8      e := (lc(hd(s)) ⊓_S lc(hd(s')), hd(s)_st, rc(hd(s)) ⊓_S rc(hd(s')));
9      c, v := augzip((z · e) : ((p · hd(s)), tl(s)) : ((p̂ · hd(ŝ)), tl(ŝ)) :
              ((p' · hd(s')), tl(s')));
10     collect := collect ∪ c;
11     avoid ∧ := 
         (v ∨ ( reasons(lc(hd(ŝ)) ⊓_C lc(hd(s')) = ⊥_C)
               ∨ reasons(rc(hd(ŝ)) ⊓_C rc(hd(s')) = ⊥_C) ));
12 if (s' ≠ ε) then
13   if (hd(s')_st ∈ (cxt(rc(p)) ∩ cxt(lc(s)))) then
14     c, v := augzip((z · hd(s')) : (p, s) : (p̂, ŝ) : ((p' · hd(s')), tl(s')));
15     collect := collect ∪ c;
16     avoid ∧ := (v ∨ ( reasons(hd(s')_st ∉ cxt(rc(p̂)))
                        ∨ reasons(hd(s')_st ∉ cxt(lc(ŝ))) ));
17 if (s = ε ∧ s' = ε) then collect := {z}, false;
18 return(collect, avoid);
```

By Lemma (9), zip($\epsilon : (\epsilon, \varphi) : (\epsilon, \varphi')$) $= \emptyset$ but zip($\epsilon : (\epsilon, \nabla_\rho(\varphi)) : (\epsilon, \varphi')$) $\neq \emptyset$. Operator $\Xi(\{\varphi\}, \{\varphi'\}, \rho)$ returns a stronger $\rho'$ s.t. zip($\epsilon : (\epsilon, \nabla_{\rho'}(\varphi)) : (\epsilon, \varphi')$) is also empty. First, we introduce the two operators reasons($q \notin cxt(cr)$), reasons($cr \sqcap_{\mathbb{C}} cr' = \bot_{\mathbb{C}}$), where $q \in Q$ and $cr, cr' \in \mathbb{C}$. The operator reasons($q \notin cxt(cr)$) returns predicate $(v_q > \kappa(cr)(q))$ if $q \notin cxt(cr)$ and $false$ otherwise. We use this operator at line (16) of the Procedure augzip. Intuitively, $cr$ is a counter prior to relaxation. Relaxation allows $q$ in the resulting context (test at line (13)). The idea is to collect possible requirements (hence disjunctions at line 16) for a resolution to forbid a meet. Here, by forbidding $q$ to belong to the relaxed context if $q$ did not belong to the context prior to relaxation. If $q$ was allowed by the context of the counter prior to relaxation, then no new resolution will forbid this by relaxing the counter. The second operator $\texttt{reasons}(cr \sqcap_{\mathbb{C}} cr' = \bot_{\mathbb{C}})$ achieves a similar result. It is used at line (11) and returns the conjunction $\{(v_q > \kappa(cr)(q))|\ (cr(q) \wedge cr'(q))$ is false$\}$. Intuitively, $cr$ is a counter prior to relaxation. The resulting counter after relaxation does meet the counter $cr'$ (test at line (7)). If $cr$ does not meet $cr'$, the operator collects the bounds that failed the meet. These will be used as possible requirements (disjunctions at line (11)) on a new resolution to ensure that after relaxation, the new counter will also not meet $cr'$, and hence fail the test at line (7). The Procedure augzip is an instrumentation of Procedure zip. Indeed, $\texttt{zip}(\epsilon : (\epsilon, \nabla_\rho(\varphi)) : (\epsilon, \varphi) : (\epsilon, \varphi'))$ mimics $\texttt{zip}(\epsilon : (\epsilon, \nabla_\rho(\varphi)) : (\epsilon, \varphi'))$. It tracks predicates on resolutions and builds an And-Or tree. Conjunctions at lines (5,11,16) reflect that no shuffle should succeed with the new relaxation. The formula $\texttt{avoid}$ will only accept resolutions that forbid the intersection.

**Lemma 10 (separation).** *Assume* $\texttt{zip}(\epsilon : (\epsilon, \varphi) : (\epsilon, \varphi'))$ *returns an empty set. If* $\texttt{zip}(\epsilon : (\epsilon, \nabla_\rho(\varphi)) : (\epsilon, \varphi) : (\epsilon, \varphi'))$ *in Procedure augzip returns the pair* $(\_, \texttt{avoid})$, *then any resolution* $\rho'$ *that satisfies* $(\texttt{avoid}[\rho'])$ *ensures* $\texttt{zip}(\epsilon : (\epsilon, \nabla_{\rho_{max}}(\varphi)) : (\epsilon, \varphi')))$ *returns the empty set, with* $\rho_{max}(q) = max(\rho(q), \rho'(q))$ *for each q in Q.*

$\Xi(\{e_1, \ldots e_n\}, \{e'_1, \ldots e'_m\}, \nabla)$ is obtained by choosing $\rho'$ to satisfy each $\texttt{avoid}_{(i,j)}$ resulting from $\texttt{zip}(\epsilon : (\epsilon, \nabla_\rho(\varphi_i)) : (\epsilon, \varphi_i) : (\epsilon, \varphi'_j))$ for $i : 1 \leq i \leq n$ and $j : 1 \leq j \leq m$. This is possible because each $\texttt{avoid}_{(i,j)}$ denotes an upward closed set of multisets.

## 7 Experimental Results

We have implemented the scheme of Figure (1) in OCaml (prototype "PCW" available from the author's homepage) and run experiments on an Intel Core 2 Duo 2.26 GHz laptop with 4GB of memory. Table (1) summarizes the results. We have considered four classical mutex algorithms, namely Burns [4], compact [6] and refined [19] versions of Szymanski's algorithm, and the related Gribomont-Zenner mutex [14] (described in appendix). The algorithms respectively appear under rows (I,II,III and IV) in Table (1). In all experiments, we used the same initial relaxation $\nabla_\rho$ for both forward and backward analysis, with $\rho(q) = 0$ for each $q$ in $Q$. For each instantiation and each algorithm, we give running times in seconds, the number of refinement steps, the number of generated counted words and the outcome of the analysis. For the last item, we write "?" to mean a trace was found by the over-approximated analysis, and write "$\sqrt{}$" to mean unreachability (i.e., safety) is established. We allocate a budget of 20 minutes for each refinement step, and write $\times$ in case the analysis exhausted the allocated time. We managed to establish mutual exclusion for the four algorithms using the backward version of the generic scheme. In forward, we could establish mutual exclusion for both algorithms (I) an (II). The analysis exhausted its time budget for the two other algorithms. Backward analysis seems to profit from the fact that it starts from an upward closed set of configurations. Forward analysis does not have that advantage.

We did experiment with non-approximated relaxations of the counters (i.e., simple accelerations). While this boosted performance, we do not report it in Table (1) since this does not strictly follow the scheme of Section (4). Combining with more systematic accelerations instead of taking one step at a time can be the subject of a natural extension of this work.

## 8 Conclusions

We have introduced a new symbolic representation for the verification of parameterized systems where processes are organized in a linear array. The new representation combines counter abstraction together with upward closure based techniques. It allows for an approximated analysis with a threshold-based precision that can be uniformly tuned. Based on the representation, we implemented a counter example based refinement scheme that illustrated the applicability and the relevance of the representation, both for forward and for backward analysis. One direction of future work is to investigate more efficient and symbolic encodings. Another direction is to investigate the applicability of such ideas, where counting constraints still converge based on a well quasi ordering argument, to other problems like parameterized systems with different topologies (trees, graphs, etc) or heap manipulating programs.

**Acknowledgments.** The author would like to thank Pierre Ganty for fruitful discussions and helpful comments on the draft.

|  |  | refine | time | steps | words | safe |
|---|---|---|---|---|---|---|
| Forward | I | 1 | .01 | 0 | 1 | ? |
|  |  | 2 | .03 | 6 | 241 | ? |
|  |  | 3 | .11 | 17 | 875 | √ |
|  | II | 1 | .01 | 0 | 1 | ? |
|  |  | 2 | .02 | 7 | 343 | ? |
|  |  | 3 | .02 | 8 | 323 | ? |
|  |  | 4 | .02 | 9 | 241 | ? |
|  |  | 5 | .15 | 11 | 297 | ? |
|  |  | 6 | .04 | 12 | 105 | ? |
|  |  | 7 | 5.85 | 171 | 5143 | √ |
|  | III | 1 | .01 | 0 | 1 | ? |
|  |  | 2 | .03 | 8 | 356 | ? |
|  |  | 3 | .04 | 10 | 406 | ? |
|  |  | 4 | .32 | 24 | 1252 | ? |
|  |  | 5 | .36 | 25 | 1248 | ? |
|  |  | 6 | .81 | 35 | 1043 | ? |
|  |  | 7 | .54 | 40 | 685 | ? |
|  |  | 8 | .04 | 12 | 149 | ? |
|  |  | 9 | 39.35 | 532 | 11006 | ? |
|  |  | 10 | >1200 | >2000 | >68000 | × |
|  | IV | 1 | .01 | 0 | 1 | ? |
|  |  | 2 | .06 | 9 | 636 | ? |
|  |  | 3 | .07 | 11 | 685 | ? |
|  |  | 4 | .07 | 12 | 602 | ? |
|  |  | 5 | .09 | 13 | 651 | ? |
|  |  | 6 | .08 | 28 | 1695 | ? |
|  |  | 7 | 2.52 | 54 | 3003 | ? |
|  |  | 8 | 10.02 | 52 | 2758 | ? |
|  |  | 9 | 3.03 | 57 | 866 | ? |
|  |  | 10 | 1.80 | 81 | 1006 | ? |
|  |  | 11 | >1200 | >2800 | >120000 | × |
| Backward | I | 1 | .02 | 2 | 151 | √ |
|  | II | 1 | .18 | 19 | 3026 | √ |
|  | III | 1 | 110.14 | 1166 | 169789 | ? |
|  |  | 2 | 158.29 | 1567 | 194425 | ? |
|  |  | 3 | 30.31 | 583 | 78942 | √ |
|  | IV | 1 | 138.10 | 932 | 233604 | ? |
|  |  | 2 | 34.33 | 434 | 129368 | √ |

**Table 1.** $\mathbb{SCW}$ based forward and backward analysis of mutex algorithms.

## References

1. P. A. Abdulla, K. Čerāns, B. Jonsson, and Y.-K. Tsay. General decidability theorems for infinite-state systems. In *Proc. LICS '96, $11^{th}$ IEEE Int. Symp. on Logic in Computer Science*, pages 313–321, 1996.
2. P. A. Abdulla, Y.-F. Chen, G. Delzanno, F. Haziza, C.-D. Hong, and A. Rezine. Constrained monotonic abstraction: A cegar for parameterized verification. In *Proc. CONCUR 2010, $21^{th}$ Int. Conf. on Concurrency Theory*, pages 86–101, 2010.
3. P. A. Abdulla, G. Delzanno, and A. Rezine. Approximated context-sensitive analysis for parameterized verification. In D. Lee, A. Lopes, and A. Poetzsch-Heffter, editors, *FMOODS/FORTE*, volume 5522 of *Lecture Notes in Computer Science*, pages 41–56. Springer, 2009.

4. P. A. Abdulla, N. B. Henda, G. Delzanno, and A. Rezine. Regular model checking without transducers (on efficient verification of parameterized systems). In *Proc. TACAS '07, $13^{th}$ Int. Conf. on Tools and Algorithms for the Construction and Analysis of Systems*, volume 4424 of *Lecture Notes in Computer Science*, pages 721–736. Springer Verlag, 2007.
5. P. A. Abdulla, B. Jonsson, M. Nilsson, and J. d'Orso. Regular model checking made simple and efficient. In *Proc. CONCUR 2002, $13^{th}$ Int. Conf. on Concurrency Theory*, volume 2421 of *Lecture Notes in Computer Science*, pages 116–130, 2002.
6. T. Arons, A. Pnueli, S. Ruah, J. Xu, and L. Zuck. Parameterized verification with automatically computed inductive assertions. In Berry, Comon, and Finkel, editors, *Proc. $13^{th}$ Int. Conf. on Computer Aided Verification*, volume 2102 of *Lecture Notes in Computer Science*, pages 221–234, 2001.
7. G. Basler, M. Mazzucchi, T. Wahl, and D. Kroening. Symbolic counter abstraction for concurrent software. In *Proceedings of CAV 2009*, volume 5643 of *LNCS*, pages 64–78. Springer, 2009.
8. B. Boigelot, A. Legay, and P. Wolper. Iterating transducers in the large. In *Proc. $15^{th}$ Int. Conf. on Computer Aided Verification*, volume 2725 of *Lecture Notes in Computer Science*, pages 223–235, 2003.
9. A. Bouajjani, P. Habermehl, and T. Vojnar. Abstract regular model checking. In *CAV04*, Lecture Notes in Computer Science, pages 372–386, Boston, July 2004. Springer Verlag.
10. E. Clarke, M. Talupur, and H. Veith. Environment abstraction for parameterized verification. In *Proc. VMCAI '06, $7^{th}$ Int. Conf. on Verification, Model Checking, and Abstract Interpretation*, volume 3855 of *Lecture Notes in Computer Science*, pages 126–141, 2006.
11. D. Dams, Y. Lakhnech, and M. Steffen. Iterating transducers. In G. Berry, H. Comon, and A. Finkel, editors, *Computer Aided Verification*, volume 2102 of *Lecture Notes in Computer Science*, 2001.
12. G. Geeraerts, J.-F. Raskin, and L. Van Begin. Expand, Enlarge and Check: new algorithms for the coverability problem of WSTS. *Journal of Computer and System Sciences*, 72(1):180–203, 2006.
13. S. M. German and A. P. Sistla. Reasoning about systems with many processes. *Journal of the ACM*, 39(3):675–735, 1992.
14. E. Gribomont and G. Zenner. Automated verification of Szymanski's algorithm. In *Proc. TACAS '98, $4^{th}$ Int. Conf. on Tools and Algorithms for the Construction and Analysis of Systems*, volume 1384 of *Lecture Notes in Computer Science*, pages 424–438, 1998.
15. T. A. Henzinger, O. Kupferman, and S. Qadeer. From pre-historic to post-modern symbolic model checking. *Formal Methods in System Design*, 23(3):303327, 2003.
16. G. Higman. Ordering by divisibility in abstract algebras. *Proc. London Math. Soc. (3)*, 2(7):326–336, 1952.
17. A. Kaiser, D. Kroening, and T. Wahl. Dynamic cutoff detection in parameterized concurrent programs. In *Proceedings of CAV*, volume 6174 of *LNCS*, pages 654–659. Springer, 2010.
18. Y. Kesten, O. Maler, M. Marcus, A. Pnueli, and E. Shahar. Symbolic model checking with rich assertional languages. In O. Grumberg, editor, *Proc. $9^{th}$ Int. Conf. on Computer Aided Verification*, volume 1254, pages 424–435, Haifa, Israel, 1997. Springer Verlag.
19. Z. Manna and A. Pnueli. An exercise in the verification of multi – process programs. In W. Feijen, A. van Gasteren, D. Gries, and J. Misra, editors, *Beauty is Our Business*, pages 289–301. Springer Verlag, 1990.
20. A. Pnueli, S. Ruah, and L. Zuck. Automatic deductive verification with invisible invariants. In *Proc. TACAS '01, $7^{th}$ Int. Conf. on Tools and Algorithms for the Construction and Analysis of Systems*, volume 2031, pages 82–97, 2001.
21. A. Pnueli, J. Xu, and L. Zuck. Liveness with (0,1,infinity)-counter abstraction. In *Proc. $14^{th}$ Int. Conf. on Computer Aided Verification*, volume 2404 of *Lecture Notes in Computer Science*, 2002.

22. T. Touili. Regular Model Checking using Widening Techniques. *Electronic Notes in Theoretical Computer Science*, 50(4), 2001. Proc. Workshop on Verification of Parametrized Systems (VEPAS'01), Crete, July, 2001.
23. M. Y. Vardi and P. Wolper. An automata-theoretic approach to automatic program verification. In *Proc. LICS '86, $1^{st}$ IEEE Int. Symp. on Logic in Computer Science*, pages 332–344, June 1986.

## A Examples

We describe in the following the mutual exclusion algorithms on which we experimented the generic scheme of Section (4) with $\mathbb{SCW}$ as symbolic representation. In all experiments, we used the same initial relaxation $\nabla_\rho$ for both forward and backward analysis, with $\rho(q) = 0$ for each $q$ in $Q$.

### A.1 Burns mutex

In this algorithm (Fig.3), local states range over $\{q_{(1:0)}, q_{(2:0)}, q_{3:0}\}$ (modeling a state where a local flag equals 0), and $\{q_{(1:1)}, q_{(3:1)}, q_{(4:1)}, q_{(5:1)}, q_{(6:1)}, q_{(7:1)}\}$ (modeling a state where a local flag equals 1). Each process interested in accessing the critical section checks twice to its left if there are other interested processes (i.e., with a flag set to 1). If there are, it returns to $q_{(1:\_)}$ (transitions $t_3$ and $t_6$). Otherwise, it continues (transitions $t_4$ and $t_9$) towards the critical section (modeled as state $q_{(6:1)}$). All processes at $q_{(5:1)}$ will successively access the critical section starting with the right most ones (transition $t_8$). Mutual exclusion is violated in case more than one process is at state $(q_{(6:1)})$.

$Q = \{q_{(1:0)}, q_{(2:0)}, q_{3:0}\} \cup \{q_{(1:1)}, q_{(3:1)}, q_{(4:1)}, q_{(5:1)}, q_{(6:1)}, q_{(7:1)}\}$ with: $T = \{t_1, \ldots t_{10}\}$ :
$t_1 : q_{(1:0)} \to q_{(2:0)}$    $t_3 : q_{(2:0)} \to q_{(1:0)} : \exists_L \{q_{(1:1)}, q_{(3:1)}, q_{(4:1)}, q_{(5:1)}, q_{(6:1)}, q_{(7:1)}\}$
$t_2 : q_{(1:1)} \to q_{(2:0)}$    $t_4 : q_{(2:0)} \to q_{(3:0)} : \forall_L \{q_{(1:0)}, q_{(2:0)}, q_{(3:0)}\}$
$t_5 : q_{(3:0)} \to q_{(4:1)}$    $t_6 : q_{(4:1)} \to q_{(1:1)} : \exists_L \{q_{(1:1)}, q_{(3:1)}, q_{(4:1)}, q_{(5:1)}, q_{(6:1)}, q_{(7:1)}\}$
$t_7 : q_{(6:1)} \to q_{(7:1)}$    $t_8 : q_{(5:1)} \to q_{(6:1)} : \forall_R \{q_{(1:0)}, q_{(2:0)}, q_{(3:0)}\}$
$t_{10} : q_{(7:1)} \to q_{(1:0)}$    $t_9 : q_{(4:1)} \to q_{(5:1)} : \forall_L \{q_{(1:0)}, q_{(2:0)}, q_{(3:0)}\}$
$S_{Init} = \{(cr_i, q_{(1:0)}, cr_i)\}$
$S_{C_F} = \{(cr_f, q_{(6:1)}, cr'_f) \cdot (cr'_f, q_{(6:1)}, cr_f)\}$

**Fig. 3.** Burns algorithm, with counters $cr_i = [v_{q_{(1:0)}} \geq 0) \wedge \wedge_{q \in Q \setminus \{q_{(1:0)}\}}(v_q = 0)]$, $cr_f = [\wedge_{q \in Q}(v_q \geq 0)]$, and $cr'_f = [(v_{q_{(6:1)}} \geq 1) \wedge \wedge_{q \in Q \setminus \{q_{(6:1)}\}}(v_q \geq 0)]$.

### A.2 Compact version of Szymanski's Mutex

This version [6] is represented in Figure (4). We flattened the original local boolean variables $s, w$ and encoded their values in process states. These range over $\{q_0, \ldots q_7\}$. The initial state is $q_0$ and $q_7$ models a process at its critical section. Processes that take transition $t_2$ are guaranteed to eventually access their critical section. At transition $t_4$, processes go to state $q_4$ where they wait for processes at state $q_1, q_2$, if any. Otherwise, transitions $t_5$ and $t_6$ are fired. Once a process is at state $q_5$, no other process can fire $t_2$, and all processes waiting at state $q_4$ can get to state $q_5$. After all processes that fired $t_2$ have gathered at state $q_5$ they can get to state $q_6$ from which they can access the critical section $q_7$ with priority to the left most processes ($t_8$).

$$
\begin{array}{l}
Q = \{q_0, q_1, \ldots q_7\} \text{ with: } T = \{t_1, \ldots t_9\} : \\
t_1 : q_0 \to q_1 \qquad\qquad\qquad\qquad t_2 : q_1 \to q_2 : \forall_{LR} \{q_0, q_1, q_2, q_4\} \\
t_3 : q_2 \to q_3 \qquad\qquad\qquad\qquad t_4 : q_3 \to q_4 : \exists_{LR} \{q_1, q_2, q_5, q_6, q_7\} \\
t_5 : q_4 \to q_5 : \exists_{LR} \{q_5, q_6, q_7\} \quad\; t_6 : q_3 \to q_5 : \forall_{LR} \{q_0, q_1, q_3, q_4\} \\
t_7 : q_5 \to q_6 : \forall_{LR} \{q_0, q_1, q_2, q_5, q_6, q_7\} \quad t_8 : q_6 \to q_7 : \forall_L \{q_0, q_1, q_2, q_4\} \\
\qquad\qquad\qquad\qquad\qquad\qquad\qquad\; t_9 : q_7 \to q_0 \\
S_{Init} = \{(cr_{init}, q_0, cr_{init})\} \\
S_{C_F} = \{(cr_{final}, q_7, cr'_{final}) \cdot (cr'_{final}, q_7, cr_{final})\}
\end{array}
$$

**Fig. 4.** Compact version of Szymanski's algorithm [6], with counters $cr_i = [(v_{q_0} \geq 0) \wedge \wedge_{q \in Q \setminus \{q_0\}}(v_q = 0)]$, $cr_f = [\wedge_{q \in Q}(v_q \geq 0)]$, and $cr'_f = [(v_{q_7} \geq 1) \wedge \wedge_{q \in Q \setminus \{q_7\}}(v_q \geq 0)]$.

### A.3 Szymanski's Algorithm

This version of Szymanski's algorithm comes from [19]. We flattened the local variable $flag$, which ranges over $\{0, 1, 2, 3, 4\}$, by encoding its value in process states. The initial state is $q_0$, and the critical section is modeled by state $q_{10}$.

$$
\begin{array}{l}
Q = \{q_0, q_1, \ldots q_{13}\} \text{ with: } T = \{t_1, \ldots t_9\} : \\
t_1 : q_0 \to q_1 \qquad t_3 : q_2 \to q_3 : \forall_{LR} \{q_0, q_1, q_2, q_3, q_7, q_8\} \\
t_2 : q_1 \to q_2 \qquad t_5 : q_4 \to q_6 : \exists_{LR} \{q_2, q_3\} \\
t_4 : q_3 \to q_4 \qquad t_7 : q_7 \to q_8 : \exists_{LR} \{q_9, q_{10}, q_{11}\} \\
t_6 : q_6 \to q_7 \qquad t_9 : q_4 \to q_5 : \forall_{LR} \{q_0, q_1, q_4, q_5, q_6, q_7, q_8, q_9, q_{10}, q_{11}\} \\
t_{10} : q_8 \to q_9 \quad t_{11} : q_9 \to q_{10} : \forall_L \{q_0, q_1, q_2, q_3\} \\
t_8 : q_5 \to q_9 \quad t_{12} : q_{10} \to q_{11} : \forall_R \{q_0, q_1, q_2, q_3, q_9, q_{10}, q_{11}\} \\
t_{13} : q_{11} \to q_0 \\
S_{Init} = \{(cr_i, q_0, cr_i)\} \\
S_{C_F} = \{(cr_f, q_{10}, cr'_f) \cdot (cr'_f, q_{10}, cr_f)\}
\end{array}
$$

**Fig. 5.** Szymanski's algorithm from [19], with counters $cr_i = (v_{q_0} \geq 0) \wedge \wedge_{q \in Q \setminus \{q_0\}}(v_q = 0)$, $cr_f = \wedge_{q \in Q}(v_q \geq 0)$, and $cr'_f = (v_{q_{10}} \geq 1) \wedge \wedge_{q \in Q \setminus \{q_{10}\}}(v_q \geq 0)$

### A.4 Griboment-Zenner Mutex

This algorithm [14] is also derived from Szymanski's algorithm (Fig.5). Its transitions are fine grained in the sense that tests and assignments are split in different atomic transitions. After encoding variable values in process states, the process states in algorithm range over the set $\{q_1, \ldots q_{13}\}$, where $q_1$ is the initial state and $q_{12}$ models a process at its critical section

```
Q = {q₁, q₁, ... q₁₃} with: T = {t₁, ... t₁₄} :
t₁ : q₁ → q₂      t₃ : q₃ → q₄ : ∀_{LR} {q₀, q₁, q₂, q₃, q₇, q₈}
t₂ : q₂ → q₃      t₅ : q₅ → q₆ : ∃_{LR} {q₂, q₃}
t₄ : q₄ → q₅      t₇ : q₇ → q₈ : ∃_{LR} {q₉, q₁₀, q₁₁}
t₆ : q₆ → q₇      t₉ : q₅ → q₉ : ∀_{LR} {q₀, q₁, q₄, q₅, q₆, q₇, q₈, q₉, q₁₀, q₁₁}
t₈ : q₈ → q₉      t₁₁ : q₁₀ → q₁₁ : ∀_L {q₀, q₁, q₂, q₃}
t₁₀ : q₉ → q₁₀    t₁₂ : q₁₁ → q₁₂ : ∀_R {q₀, q₁, q₂, q₃, q₉, q₁₀, q₁₁}
t₁₃ : q₁₂ → q₁₂   t₁₄ : q₁₃ → q₁
S_{Init} = {(cr_i, q₁, cr_i)}
S_{C_F} = {(cr_f, q₁₂, cr'_f) · (cr'_f, q₁₂, cr_f)}
```

**Fig. 6.** Gribomont-Zenner algorithm [14], with counters $cr_i = [(v_{q_1} \geq 0) \wedge \wedge_{q \in Q \setminus \{q_1\}}(v_q = 0)]$, $cr_f = [\wedge_{q \in Q}(v_q \geq 0)]$, and $cr'_f = [(v_{q_{12}} \geq 1) \wedge \wedge_{q \in Q \setminus \{q_{12}\}}(v_q \geq 0)]$

## B  Proofs

### B.1  Section 4

**Lemma 1 (reachability).** *Algorithm (1) always terminates. If it returns* `unreachable`, *then* $(Init \xrightarrow{*} C_F)$ *does not hold for the parameterized system* $\mathcal{P} = (Q, T)$. *Otherwise, it returns a trace* $e_0 \cdot t_1 \cdot e_1 \cdots e_m$ *with* $e_0 \in E_{Init}$, $\{e_m\} \sqcap_\mathbb{S} E_{C_F} \neq \emptyset$, *and* $e_{i+1} \in \nabla(post_{t_{i+1}}(e_i))$ *for each* $i : 0 \leq i < m$.

*Proof.* Let $\mathcal{W}_k$ and $\mathcal{V}_k$ be the sets $\mathcal{W}$ and $\mathcal{V}$ obtained at (line (2)) at the $k^{th}$ iteration of the loop. We can show the following four propositions by induction on $k$:

a) for each $(e, \tau)$ in $\mathcal{V}_k \cup \mathcal{W}_k$, $\tau$ equals $e_0 \cdot t_1 \cdots e_n$ for some $n \leq k$, with $e_0 \in E_{Init}$, and $e_{i+1} \in \nabla(post_{t_{i+1}}(e_i))$ for each $i : 0 \leq i < n$
b) *Init* is subset of $[\![\mathcal{V}_k]\!] \cup [\![\mathcal{W}_k]\!]$, and $\{c' | c \longrightarrow_t c' \text{ for some } c \in [\![\mathcal{V}_k]\!], t \in T\}$ (i.e., the successors of $[\![\mathcal{V}_k]\!]$) are in $[\![\mathcal{W}_k]\!] \cup [\![\mathcal{V}_k]\!]$.
c) for each $e$ in $\mathcal{V}$, $\{e\} \sqcap_\mathbb{S} E_{C_F}$ is empty
d) the set $\{e | (e, \tau) \in \mathcal{V}_k \cup \mathcal{W}_k\}$ is minimal wrt $\sqsubseteq_\mathbb{S}$, i.e., for any $e, e'$, neither $e \sqsubseteq_\mathbb{S} e'$ nor $e' \sqsubseteq_\mathbb{S} e$ holds.

Base case: $k = 0$. Propositions (a) and (c) are verified. Requirement (1) ensures proposition (b). For proposition (d), $E_{Init}$ is assumed to be minimal (otherwise choose any minimal subset of it).

Suppose the propositions hold up to $k$. We show they hold for $k + 1$. Elements added to $\mathcal{V}$ need to fail the test at (line (4)). So proposition (c) holds. The test at (line (10)) and the conditions at (lines (11,12)) guarantee proposition (d). The form of the added tuple at (line (12)) ensures proposition (a).

For proposition (b), we show that i) $[\![\mathcal{V}_k \cup \mathcal{W}_k]\!] \subseteq [\![\mathcal{V}_{k+1} \cup \mathcal{W}_{k+1}]\!]$, and that ii) the successors of $[\![e_c]\!]$ are also in $[\![\mathcal{V}_{k+1} \cup \mathcal{W}_{k+1}]\!]$. The only modifications to $\mathcal{V}_k \cup \mathcal{W}_k$ occur at (lines (11,12)). If $e'$ is removed at (lines (11,12)) from $\mathcal{V} \cup \mathcal{W}$, then the conditions at (lines (11,12)) ensure that $e \sqsubseteq_\mathbb{S} e'$ for some $e \in New_t$. The element $e$ is added to $\mathcal{V} \cup \mathcal{W}$, hence $[\![\mathcal{V}_k \cup \mathcal{W}_k]\!] \subseteq [\![\mathcal{V}_{k+1} \cup \mathcal{W}_{k+1}]\!]$. For condition (ii), observe that requirements (3.a) and (6) ensure that if $c \in [\![e_c]\!]$ and $c \longrightarrow_t c'$, then there is $e' \in New_t$ with $c' \in [\![e']\!]$. The conditions at (lines (10,11,12)) ensure that $e'$ is added to $\mathcal{V}$ unless it entails an element in $\mathcal{V} \cup \mathcal{W}$.

*Partial correctness.* Suppose the Algorithm returns `unreachable`. Then $\mathcal{W}$ was empty at some iteration. Combined with proposition (b), we get that $[\![\mathcal{V}]\!]$ is a fixpoint that includes all reachable configurations. Proposition (c) ensures that $[\![\mathcal{V}]\!] \cap C_F$ is empty. If the algorithm returns $(\texttt{reach}, \tau)$, then the test at (line (4)) ensures that $\{e_c\} \sqcap_\mathbb{S} E_{C_F}$ is non empty. Moreover, (line (3)) together with proposition (a) ensure that $\tau = e_0 \cdot t_1 \cdot e_1 \cdots e_m$ satisfies $e_0 \in E_{Init}$, $e_m = e_c$, and $e_{i+1} \in \nabla(\text{post}_{t_{i+1}}(e_i))$ for each $i : 0 \leq i < m$.

*Termination.* Suppose the algorithm does not terminate. That means we add an infinite number of elements to $\mathcal{V}$. Consider a sequence $(e_1, e_2, \ldots)$ where each $e_k$ is some element added at iteration $k$. Proposition (d) and transitivity guarantee that in this sequence, $i < j$ implies that $e_i \not\sqsubseteq_\mathbb{S} e_j$. Indeed, if $e_i \sqsubseteq_\mathbb{S} e_j$ for some $i < j$, then $e_i \notin \mathcal{V}_j$ when $e_j$ was added at (line (11)) to $\mathcal{V}_j$. This means that $e_i$ was removed by some element $e_k$ added to $\mathcal{V}_k$ at a later iteration $k : i < k < j$ such that $e_k \sqsubseteq_\mathbb{S} e_i$. By repeating this reasoning, and transitivity of $\sqsubseteq_\mathbb{S}$, we deduce that $\mathcal{V}_{j-1}$ had an element $e_{j-1}$ such that $e_{j-1} \sqsubseteq_\mathbb{S} e_j$. Thus, $e_j$ should not have passed the test at (line (10)), and hence not been added to $\mathcal{V}_j$. The existence of such an infinite sequence contradicts requirement (3.b) since, according to the computation at (line (8)), all elements in $\mathcal{V}$ are of the form $\nabla(e)$ for some element $e$.

**Lemma 2 (Refinement).** *Given a $\nabla$-trace $\tau = e_0 \cdot t_1 \cdots e_m$ with $\{e_m\} \sqcap_\mathbb{S} E_{C_F} \neq \emptyset$, Algorithm (2) terminates. If it returns $(\texttt{reach}, \tau)$, then there are $c_0, \ldots c_m$ in $C$, with $c_0 \in Init$, $c_m \in C_F$ and s.t. $c_i \longrightarrow_{t_{i+1}} c_{i+1}$ for $i : 0 \leq i < m$. Otherwise, it returns a stronger relaxation $\nabla'$ such that no relaxation $\nabla''$ that is stronger than $\nabla'$ can have a $\nabla''$-trace $e'_0 \cdot t_1 \cdots e'_m$ with $\{e'_m\} \sqcap_\mathbb{S} E_{C_F} \neq \emptyset$ and where $e_i \sqsubseteq_\mathbb{S} e'_i$ for $i : 0 \leq i \leq m$.*

*Proof.* Termination is guaranteed by the fact that the number of iterations of the loop at (line (2)) is bounded by the finite size of the trace $\tau$, and that the other operations are effective. The algorithm returns $(\texttt{reach}, \tau)$ if it succeeds in building a sequence $(Current_m \cdot t_{m-1} \cdots Current_0)$ with $Current_m = \{e_m\} \sqcap_\mathbb{S} E_{C_F}$, and $Current_i = \{e_i\} \sqcap_\mathbb{S} (\bigcup_{e \in Current_{i+1}} \text{pre}_{t_{i+1}}(e))$ for each $i : 0 \leq i < m$, and $Current_0 \sqcap_\mathbb{S} E_{Init} \neq \emptyset$. Requirements (2), (5) and (6) on $\text{pre}_t(.)$ and on $\sqcap_\mathbb{S}$ guarantee the existence of configurations $c_0, \ldots c_m$ with $c_0$ in $[\![Current_0]\!]$, and s.t. $c_i \longrightarrow_{t_{i+1}} c_{i+1}$ and $c_{i+1}$ in $[\![Current_{i+1}]\!]$ for each $i : 0 \leq i < m$. Suppose the Algorithm returns instead a relaxation operator $\nabla'$ such that there exists a $\nabla''$-trace $(e'_0 \cdot t_1 \cdots e'_m)$, for some $\nabla''$ stronger than $\nabla'$, with $e_0 \sqsubseteq_\mathbb{S} e'_0$, $\{e'_m\} \sqcap_\mathbb{S} E_{C_F} \neq \emptyset$, and $e'_{i+1} \in \nabla''(\text{post}_{t_{i+1}}(e'_i))$ with $e_{i+1} \sqsubseteq_\mathbb{S} e'_{i+1}$ for each $i : 0 \leq i < m$. This would happen at line (5) or at line (11). If at line (5), then let $Current_m, Current_{m-1}, \ldots$ be the sequence of sets $Current$ manipulated at each iteration. In a similar manner, build the sequence $Current'_m, Current'_{m-1}, \ldots$ that we would obtain from the $\nabla''$-trace $(e'_0 \cdot t_1 \cdots e'_m)$. Requirements (1) and (6.b) guarantee that $Current_j \sqsubseteq_\mathbb{S} Current'_j$ for each $j : 0 \leq j \leq m$. The fact that Algorithm (2) returned at line (5) says that there is a $i : 0 \leq i < m$ s.t. $\nabla' = \Xi(\text{post}_{t_i}(e_i), Current_{i+1}, \nabla)$ with $(\nabla'(\text{post}_{t_i}(e_i)) \sqcap_\mathbb{S} Current_{i+1} = \emptyset)$. Since $e_i \sqsubseteq_\mathbb{S} e'_i$ we deduce (requirement (6.b)) that $\text{post}_{t_i}(e_i) \sqsubseteq_\mathbb{S} \text{post}_{t_i}(e'_i)$, and hence (requirement (3.b)) that $\nabla''(\text{post}_{t_i}(e_i)) \sqsubseteq_\mathbb{S} \nabla''(\text{post}_{t_i}(e'_i))$. Since $Current_{i+1} \sqsubseteq_\mathbb{S} Current'_{i+1}$, we deduce $(\nabla''(\text{post}_{t_i}(e'_i)) \sqcap_\mathbb{S} Current'_{i+1} = \emptyset)$ as otherwise $\nabla'(\text{post}_{t_i}(e_i)) \sqcap_\mathbb{S} Current_{i+1} \neq \emptyset$. If at line (11), then $\nabla' = \Xi(\{e_0\}, E_{Init}, \nabla)$, with $\nabla'(e_0) \sqcap_\mathbb{S} E_{Init} = \emptyset$. Since $e_0 \sqsubseteq_\mathbb{S} e'_0$, we deduce $\nabla'(e_0) \sqsubseteq_\mathbb{S} \nabla''(e'_0)$ and $\{\nabla''(e'_m)\} \sqcap_\mathbb{S} E_{Init} = \emptyset$. Hence $\tau'$ would have $\{e'_m\} \sqcap_\mathbb{S} E_{C_F} = \emptyset$, which contradicts its definition.

**Theorem 1.** *Each iteration of the Algorithm depicted in Figure (1) terminates. If it returns* `unreachable`, *then* $Init \xrightarrow{*} C_F$ *does not hold. If it returns* (`reachable`, $\tau$), *then* $Init \xrightarrow{*} C_F$ *via the transitions in $\tau$. In addition, a false positive cannot be generated twice.*

*Proof.* From Lemma (1) and Lemma (2) and the requirement that $\Xi$ returns a stronger relaxation operator.

### B.2 Section 5

**Lemma 3.** *For any $k \in \mathbb{N}$, $(\mathbb{C}_k, \sqsubseteq_\mathbb{C})$ is a well quasi ordering. In fact, from every infinite sequence $(cr_1, cr_2, \ldots)$, we can extract an infinite subsequence $(cr_{i_1} \sqsubseteq_\mathbb{C} cr_{i_2} \sqsubseteq_\mathbb{C} \ldots)$.*

*Proof.* Let $(cr_1, cr_2, \ldots)$ be an infinite sequence of counters. Fix a state $q$. If the number of counters for which $cr_m(q) = (v_q \geq b)$ (regardless of $b$) is infinite, then remove from the sequence all the counters for which $cr_m(q) = (v_q = b')$ for some $b'$. Otherwise, if the number of counters for which $cr_m(q) = (v_q \geq b)$ (regardless of $b$) is finite, then there is a $b_0 < k$ such that the number of counters for which $cr_m(q) = (v_q = b_0)$ is infinite. Keep those counters and remove all others from the sequence. By repeating this procedure for each state $q$ in $Q$, we obtain a new infinite sequence of counters $(cr_{m_1}, cr_{m_2}, \ldots)$ for which, for any $b$, $cr_{m_a}(q) = (v_q = b)$ iff $cr_{m_{a'}}(q) = (v_q = b)$. Fix a state $q$ for which $cr_{m_1}(q) = (v_q \geq b)$. It is possible to extract from the sequence another infinite sequence $(cr_{n_1}, cr_{n_2}, \ldots)$ such that if $cr_{n_a}(q) = (v_q \geq b_{n_a})$ and $cr_{n_{a'}}(q) = (v_q \geq b_{n_{a'}})$ with $n_a < n_{a'}$, then $b_{n_a} \leq b_{n_{a'}}$. By repeating this for each state $q$, we obtain an infinite sequence in which $(cr_{i_1} \sqsubseteq_\mathbb{CW} cr_{i_2} \sqsubseteq_\mathbb{CW} \ldots)$.

**Lemma 4 (Well formedness).** *Assume a counted word $\varphi = (l_1, q_1, r_1) \cdots (l_n, q_n, r_n)$. For each $i : 1 \leq i \leq n$, let $p_i^r = ((\underline{\varphi})_{[1,i-1]})^\#$ and $p_i^l = ((\underline{\varphi})_{[i+1,n]})^\#$. Then:*

- *Each $l_i(q)$ equals $(v_q = (p_i^r(q)))$ or $(v_q \geq k)$ for some $k$ in $\{0, \ldots (p_i^r(q))\}$.*
- *Each $r_i(q)$ equals $(v_q = (p_i^l(q)))$ or $(v_q \geq k)$ for some $k$ in $\{0, \ldots (p_i^l(q))\}$.*

*Proof.* Well formedness requires, for each $i : 1 \leq i \leq n$, that $l_i[(\underline{\varphi})_{[1,i-1]}]$ and $r_i[(\underline{\varphi})_{[i+1,n]}]$. The rest follows from the allowed predicates in the counters.

**Lemma 5 (Entailment).** *The relation $\sqsubseteq_\mathbb{CW}$ is both reflexive and transitive. In addition, $\varphi \sqsubseteq_\mathbb{CW} \varphi'$ implies $[\![\varphi']\!] \subseteq [\![\varphi]\!]$.*

*Proof.* Follows from reflexivity and transitivity of $\sqsubseteq_{\mathbb{C}}$. Also, if $w \models^h \varphi'$ and $\varphi \sqsubseteq^{h'} \varphi'$ then $w \models^{h \circ h'} \varphi$ with $\circ$ is function composition.

**Theorem 2 (WQO).** *For any fixed $k \in \mathbb{N}$, the pair $(\mathbb{CW}_k, \sqsubseteq)$ is a well quasi ordering.*

*Proof.* Higman's Lemma [16] states that if $(\Sigma, \preceq)$ is a well quasi ordering, then the pair $(\Sigma^*, \preceq^*)$ is also a well quasi ordering[9]. We let $\Sigma = \mathbb{C}_k \times Q \times \mathbb{C}_k$, and $(l, q, r) \preceq (l', q', r')$ if $l \sqsubseteq_{\mathbb{C}} l'$ and $q = q'$ and $r \sqsubseteq_{\mathbb{C}} r'$. Observe that $\mathbb{CW}_k \subseteq \Sigma^*$, and that $\preceq^*$ coincides with $\sqsubseteq_{\mathbb{CW}}$. Hence, showing that $(\Sigma, \preceq)$ is a well quasi ordering establishes the result. Lemma (3) states that $(\mathbb{C}_k, \sqsubseteq_{\mathbb{C}})$ is a well quasi ordering and that from every infinite sequence $(cr_1, cr_2, \ldots)$, we can extract an infinite subsequence $(cr_{i_1} \sqsubseteq_{\mathbb{C}} cr_{i_2} \sqsubseteq_{\mathbb{C}} \ldots)$. Given an infinite sequence $(l_1, q_1, r_1), (l_2, q_2, r_2), \ldots$ we can extract an infinite sequence $((l_{m_1}, q_{m_1}, r_{m_1}), (l_{m_2}, q_{m_2}, r_{m_2}), \ldots)$ in which $q_{m_a} = q_{m_b}$ for all $a \neq b$. We use Lemma (3) to deduce the existence of an infinite sequence $((l_{n_1}, q_{n_1}, r_{n_1}) \preceq (l_{n_2}, q_{n_2}, r_{n_2}) \preceq \ldots)$.

**Lemma 6 (Strengthening).** *Given a well formed word $\varphi$, the strengthening procedure terminates and yields a unique well formed word $\varphi'$ s.t. $[\![\varphi]\!] = [\![\varphi']\!]$ and $\varphi \sqsubseteq_{\mathbb{CW}} \varphi'$.*

*Proof.* Sketch. The rules of Figure (5) only strengthen the counters. Hence, $\varphi \sqsubseteq_{\mathbb{CW}} \varphi'$ and $[\![\varphi]\!] \supseteq [\![\varphi']\!]$. We first show that the rules of Figure (5) preserve denotation. We describe the cases `left` and `right`. The two other rules are symmetric. Suppose the counted word $\varphi$ can be written both as the concatenation $\varphi_p \cdot (l, q, r) \cdot (l', q', r') \cdot \varphi_s$ and as the concatenation $q_1 \cdot q_2 \cdots q_n$, with $i = |\varphi_p|$. If $w$ in $Q^*$ verifies $w \models^h \varphi$, then: $w_{[h(i+1)]} = q$, $w_{[h(i+2)]} = q'$, $l(w_{[1, h(i+1)-1]})$, $r(w_{[h(i+1)+1, n]})$, $l'(w_{[1, h(i+2)-1]})$, and $r'(w_{[h(i+2)+1, n]})$. Well formedness of $\varphi$ and the allowed predicates in the counters (Lemma (4)) ensure that $l'(w_{[1, h(i+1)]})$. Therefore, $l' \ominus_{\mathbb{C}} cr_q(w_{[1, h(i+1)-1]})$. So both $l$ and $l'$ accept $w_{[1, h(i+1)-1]}$ and rule `left` does not affect denotation. In addition, $r' \oplus_{\mathbb{C}} cr_{q'}$ accepts $(w_{[h(i+2), n]})$ and $l' \ominus_{\mathbb{C}} (l \oplus_{\mathbb{C}} cr_q)$ accepts $(w_{[h(i+1)+1, h(i+2)-1]})$. So $r' \oplus_{\mathbb{C}} cr_{q'} l' \ominus_{\mathbb{C}} (l \oplus_{\mathbb{C}} cr_q)$ accepts $(w_{[h(i+1)+1, h(i+2)-1]}) \cdot (w_{[h(i+2), n]})$. As a result, both $r$ and $r' \oplus_{\mathbb{C}} cr_{q'} l' \ominus_{\mathbb{C}} (l \oplus_{\mathbb{C}} cr_q)$ accept $(w_{[h(i+1)+1, h(i+2)-1]}) \cdot (w_{[h(i+2), n]})$ and rule `right` does not affect the denotation. Observe that with a similar reasoning, we get that the rules preserve well formedness. By induction we get that $[\![\varphi]\!] \subseteq [\![\varphi']\!]$. Termination can be obtained as follows. At each rule, manipulated and obtained counted words are well formed. Using Lemma (4), we deduce that all counters belong to a finite lattice in which rules are monotonic functions that strengthen a counter and keep the others unchanged. Unicity can be obtained by contradiction. Suppose two different counted words are obtained as strengthenings of the same well formed counted word. The words can only differ in their counters. Pick different corresponding counters. Given the allowed forms for the predicates (Lemma (4)), we deduce that at least one predicate associated to some state is strictly stronger in one of the counters. If we apply to the word with a weaker predicate, the sequence of rules that were applied to the word with a stronger predicate, we would get a strictly stronger predicate. This contradicts having reached a fixpoint for the counted word with a weaker predicate.

---

[9] With $\sigma_1 \cdots \sigma_n \preceq^* \sigma'_1 \cdots \sigma'_m$ iff there is a strictly increasing $h : \overline{n} \to \overline{m}$ with $\sigma_i \preceq \sigma'_{h(i)}$

## B.3 Section 6

**Lemma 7 (Relaxation).** *Given $\varphi$ in $\mathbb{CW}$ and resolutions $\rho \subseteq \rho'$, we have: $\nabla_\rho(\varphi) \sqsubseteq_{\mathbb{CW}} \nabla_{\rho'}(\varphi) \sqsubseteq_{\mathbb{CW}} \varphi$ and $\nabla_\rho(\varphi)$ is in $\mathbb{SCW}_{max(0,2k-1)}$ with $k = max\{\rho(q)|\ q\ in\ Q\}$.*

*Proof.* Sketch. Without strengthening, $k$ is the highest precision for the counters in $\rho \subseteq \rho$, and the lemma clearly holds. Using a similar reasoning to the one used for proving unicity of the strengthening result, we can show that $\nabla_\rho(\varphi) \sqsubseteq_{\mathbb{CW}} \nabla_{\rho'}(\varphi) \sqsubseteq_{\mathbb{CW}} \varphi$. Indeed, if for example $\nabla_\rho(\varphi) \not\sqsubseteq_{\mathbb{CW}} \nabla_{\rho'}(\varphi)$, then there is a counter $cr'$ in $\nabla_{\rho'}(\varphi)$ that does not entail a corresponding counter $cr$ in $\nabla_\rho(\varphi)$, i.e., $cr \not\sqsubseteq_{\mathbb{C}} cr'$. This is not possible. Indeed, before strengthening, $\nabla_\rho(\varphi)$ and $\nabla_{\rho'}(\varphi)$ are both well formed with the same base and strengthening in $\nabla_{\rho'}(\varphi)$ starts with stronger counters than the corresponding ones in $\nabla_\rho(\varphi)$. By applying to $\nabla_{\rho'}(\varphi)$ the sequence of strengthening rules used to strengthen $\nabla_\rho(\varphi)$, we obtain a $cr'$ that is at least as strong as $cr$. In addition, strengthening cannot introduce arbitrary precision. The strongest precision $2k - 1$ (for $k \geq 1$) derived by strengthening is obtained when both left and right counters in some tuple $(l, q, r)$ associate the predicate $v_q = (k-1)$ with the state $q$. In fact, one can show by induction on the number of applications of the strengthening rules (of $\nabla_\rho(\varphi)$), that for any state $q'$, $\kappa(l \oplus_{\mathbb{C}} cr_q \oplus_{\mathbb{C}} r)(q') \leq max(0, 2k - 1)$ is an invariant for each tuple $(l, q, r)$.

**Lemma 8 (Post and Pre).** *Given a strengthened word $\varphi$ and a transition $t$ we can compute a set of words $post_t(\varphi)$ (resp. $pre_t(\varphi)$) such that $post_t(\varphi) \sqsubseteq_{\mathbb{CW}} post_t(\varphi')$ (resp. $pre_t(\varphi) \sqsubseteq_{\mathbb{CW}} pre_t(\varphi')$) if $\varphi \sqsubseteq_{\mathbb{CW}} \varphi'$. In addition, $[\![post_t(\varphi)]\!]$ and $[\![pre_t(\varphi)]\!]$ respectively equal $\{c'|\ c \longrightarrow_t c'\ with\ c\ in\ [\![\varphi]\!]\}$, and $\{c'|\ c' \longrightarrow_t c\ with\ c\ in\ [\![\varphi]\!]\}$.*

*Proof.* The construction is given in Section (6).

*Meet of two counted words.* First, we introduce a number of notations. Let $u, v, u', v'$ be counted words in $\mathbb{CW}$. We write $h_u^v$, to mean a strictly increasing injection from $\overline{|u|}$ to $\overline{|v|}$. Given two injections $h_u^v$ and $h_{u'}^{v'}$, we also write $h_{u\bullet u'}^{v \bullet v'}$ to mean the mapping that sends $i$ to $h_u^v(i)$ if $i \in \{1, \ldots |u|\}$ and to $h_{u'}^{v'}(i-|u|)+|v|$ if $i \in \{|u|+1, \ldots |u|+|u'|\}$. We will make use of Definition (1) in order to prove partial correctness (Lemmata (11,9)). Definition (1) describes conditions for a tuple $(w, (u : v), (u', v'))$ to be good wrt to injections $h_u^w$ and $h_{u'}^w$. Roughly, if a tuple $(w, (u : v), (u', v'))$ is good wrt $h_u^w$ and $h_{u'}^w$ then $w \models^{h_u^w} u$ and $w \models^{h_{u'}^w} u'$. Moreover, if such a good tuple is supplied to the zip procedure, then all recursive calls will have as arguments good tuples whose associated injections extend (in a sens that will be made clear in Lemma (11)) $h_u^w$ and $h_{u'}^w$.

**Definition 1 (Goodness).** *Given counted words $w, u, v, u', v'$ and injections $h_u^w, h_{u'}^w$, we say the tuple $(w : (u, v) : (u', v'))$ is $(h_u^w, h_{u'}^w)$-good, if $h_u^w(\overline{|u|}) \cup h_{u'}^w(\overline{|u'|}) = \overline{|w|}$ and the following holds for each $j : 1 \leq j \leq |w|$:*

1. *if $h_u^w(i) = j \notin h_{u'}^w(\overline{|u'|})$. Define $i' = max\{0\} \cup \{k|\ h_{u'}^w(k) \leq j\}$. Then:*
   (a) *$(u_i)_{st} \in cxt(rc(u'_{[1,i']})) \cap cxt(lc(u'_{[i'+1,|u'|]} \cdot v'))$, and*
   (b) *$w_j = u_i$.*
2. *if $h_u^w(i) = h_{u'}^w(i') = j$, then:*

(a) $lc(u_i) \sqcap_{\mathbb{C}} lc(u'_{i'}) \neq \bot_{\mathbb{C}}$ and $rc(u_i) \sqcap_{\mathbb{C}} rc(u'_{i'}) \neq \bot_{\mathbb{C}}$, and

(b) $w_j = (lc(u_i) \sqcap_{\mathbb{C}} lc(u'_{i'}), (u_i)_{st}, rc(u_i) \sqcap_{\mathbb{C}} rc(u'_{i'}))$.

3. if $h^w_{u'}(i') = j \notin h^w_u(\overline{|u|})$, similar to the first case with $i, u, v, i', u', v'$ respectively replaced by $i', u', v', i, u, v$.

Lemma (11) establishes that given a good tuple $(z : (p, s) : (p', s'))$ as argument, the procedure zip computes all counted words that entail $(p \cdot s)$ and $(p' \cdot s')$ and that have $z$ as prefix.

**Lemma 11 (zip correctness).** *Given an $(h^z_p, h^z_{p'})$-good tuple $(z : (p, s) : (p', s'))$, the procedure* $\texttt{zip}(z : (p, s) : (p', s'))$ *computes all counted words* $z \cdot z'$ *such that* $(z \cdot z' : (p \cdot s, \epsilon) : (p' \cdot s', \epsilon))$ *is* $(h^{z \bullet z'}_{p \bullet s}, h^{z \bullet z'}_{p' \bullet s'})$-*good*.

*Proof.* Sketch. For termination, use $|s| + |s'|$ as a ranking function. For partial correctness, proceed by induction on $|s| + |s'|$ to show that if claim (5) holds then claim (6) also holds, where claims (5) and (6) are given by:

$$(z : (p, s) : (p', s')) \text{ is } (h^z_p, h^z_{p'})\text{-good}. \qquad (5)$$

$$\texttt{zip}(z : (p, s) : (p', s')) = \left\{ z \cdot z' \mid (z \cdot z' : (p \cdot s, \epsilon) : (p' \cdot s', \epsilon)) \text{ is } (h^{z \bullet z'}_{p \bullet s}, h^{z \bullet z'}_{p' \bullet s'})\text{-good} \right\} \qquad (6)$$

Base case: assume $|s| + |s'| = 0$. Among the guards at lines (2,5,9,12) only the guard at line (12) is enabled. The procedure $\texttt{zip}(z : (p, \epsilon) : (p', \epsilon))$ returns $\{z\}$ at line (13) for which $(z : (p, \epsilon) : (p', \epsilon))$ is $(h^z_p, h^z_{p'})$-good by assumption.

Induction: suppose Lemma 11 holds for all $(h^z_p, h^z_{p'})$-good tuples $(z : (p, s) : (p', s'))$ with $|s| + |s'|$ up to $n_0$. Assume claim (5) is true with $|s| + |s'| = (n_0 + 1)$. We show claim (6) holds. We first consider the case where both $s$ and $s'$ are non empty.

⊆) Suppose $z'' \in \texttt{zip}(z : (p, s) : (p', s'))$. By assumption, among the guards at lines (2, 5, 9, 12), only the last one is disabled. By definition of the Procedure (zip), $z''$ has to be obtained from one of the following calls to Procedure (zip):

i) The call to Procedure (zip) at line (4). We describe why it is the case that $(z \cdot hd(s) : (p \cdot hd(s), tl(s)) : (p', s'))$ is $(h^{z \bullet hd(s)}_{p \bullet hd(s)}, h^{z \cdot hd(s)}_{p'})$-good. For each $j : 1 \leq j \leq |z|$, conditions of items 1,2,3 in Definition (1) are guaranteed by claim (5). For $j = |z| + 1$, item (1) in Definition (1) requires $(u_i)_{st} \in cxt(rc(u'_{[1,i']})) \cap cxt(lc(u'_{[i'+1,|u'|]} \cdot v'))$, and $w_j = u_i$. This is guaranteed by the condition at line (3) which ensures that $hd(s)$, $p'$ and $s'$ (respectively playing the roles of $w_j$, $u'_{[1,i']}$, and $u'_{[i'+1,|u'|]} \cdot v'$) satisfy $(hd(s))_{st} \in (cxt(rc(p')) \cap cxt(lc(s')))$. We can apply the induction hypothesis since $|tl(s)| + |s'| = n_0$ and obtain that $z'' = z \cdot hd(s) \cdot z'''$ with $(z'' : (p \cdot s, \epsilon) : (p' \cdot s', \epsilon))$ is $(h^{z \bullet hd(s) \bullet z'''}_{p \bullet hd(s) \bullet tl(s)}, h^{z \cdot hd(s) \bullet z'''}_{p' \bullet s'})$-good.

ii) The call to Procedure (zip) at line (8). We explain why it is the case that $(z \cdot e : (p \cdot hd(s), tl(s)) : (p' \cdot hd(s'), tl(s')))$ is $(h^{z \bullet e}_{p \bullet hd(s)}, h^{z \bullet e}_{p' \bullet hd(s')})$-good. For each $j : 1 \leq j \leq |z|$, conditions of items 1,2,3 in Definition (1) are guaranteed by claim (5). For $j = |z|+1$, item (2) in Definition (1) requires that both $lc(u_i) \sqcap_{\mathbb{C}} lc(u'_{i'})$ and $rc(u_i) \sqcap_{\mathbb{C}} rc(u'_{i'})$ are different from $\bot_{\mathbb{C}}$. In addition, it requires that $w_j = (lc(u_i) \sqcap_{\mathbb{C}} lc(u'_{i'}), (u_i)_{st}, rc(u_i) \sqcap_{\mathbb{C}} rc(u'_{i'}))$. This is guaranteed by the condition at line (6) that ensures that $(lc(hd(s)) \sqcap_\mathbb{C} lc(hd(s'))) \neq \bot_\mathbb{C})$ and $(hd(s)_{st} = hd(s')_{st})$ and $(rc(hd(s)) \sqcap_\mathbb{C} rc(c') \neq \bot_\mathbb{C})$ and the assignment at line (7) which ensures that

$$e = (lc(hd(s)) \sqcap_\mathbb{C} lc(hd(s')), hd(s)_{st}, rc(hd(s)) \sqcap_\mathbb{C} rc(hd(s)'))$$

Here, $e$ plays the role of $w_j$, $hd(s)$ the one of $u_i$, and $hd(s')$ the one of $u'_{i'}$. We can apply the induction hypothesis since $|tl(s)| + |tl(s')| = n_0 - 1$ and obtain that $z'' = z \cdot e \cdot z'''$ with $(z'' : (p \cdot s, \epsilon) : (p' \cdot s', \epsilon))$ is $(h^{z \bullet hd(s) \bullet z'''}_{p \bullet hd(s) \bullet tl(s)}, h^{z \cdot hd(s) \bullet z'''}_{p' \bullet s'})$-good.

- The call at line (11). Replace $p$ and $s$ by respectively $p'$ and $s'$ in the first case.

⊇) Consider a $z \cdot z'$ where $(z \cdot z' : (p \cdot s, \epsilon) : (p' \cdot s', \epsilon))$ is $(h^{z \bullet z'}_{p \bullet s}, h^{z \bullet z'}_{p' \bullet s'})$-good and $(z : (p,s) : (p', s'))$ is $(h^z_p, h^z_{p'})$-good. We want to show that $z \cdot z'$ is among the values returned by $\text{zip}(z : (p,s) : (p', s'))$. Definition (1) guarantees that: $\overline{|z \cdot z'|} = h^{z \bullet z'}_{p \bullet s}(\overline{|p \cdot s|}) \cup h^{z \bullet z'}_{p' \bullet s'}(\overline{|p' \cdot s'|})$ and $\overline{|z|} = h^z_p(\overline{|p|}) \cup h^z_{p'}(\overline{|p'|})$. Hence, the word $z'$ is not empty since $s \cdot s' \neq \epsilon$. Let $j = |z| + 1$, we have $(z \cdot z')_j = z'_1$. We know $j \in h^{z \bullet z'}_{p \bullet s}(\overline{|p \cdot s|}) \cup h^{z \bullet z'}_{p' \bullet s'}(\overline{|p' \cdot s'|})$. Since $\overline{|z|} = h^z_p(\overline{|p|}) \cup h^z_{p'}(\overline{|p'|})$, there are three cases:

  i) $j = h^{z \bullet z'}_{p \bullet s}(|p| + 1)$ and $h^{z \bullet z'}_{p' \bullet s'}(|p'|) < j < h^{z \bullet z'}_{p' \bullet s'}(|p'| + 1)$. This corresponds to the item (1) in Definition (1). Here, $z'_1 = (z \cdot z')_{|z|+1}$ is playing the role of $w_j$, $(p \cdot s)_{|p|+1} = hd(s)$ the one of $u_i$, $|p'|$ the one of $i'$, $p \cdot s$ the one of $u$, $p' \cdot s'$ the one of $u'$, and $\epsilon$ the one of $v$ and $v'$. As a result, the condition (2) of Definition (1) which states that $(u_j)_{st} \in cxt(rc(u'_{[1,i']})) \cap cxt(lc(u'_{[i'+1,|u'|]} \cdot v'))$ guarantees that the guard at line (3) will be satisfied. $\text{zip}(z : (p,s) : (p', s'))$ will therefore call $\text{zip}(z \cdot hd(s) : (p \cdot hd(s), tl(s)) : (p', s'))$ which satisfies the induction hypothesis (since $|tl(s)| + |s'| = n_0$) and returns all $(h^{z \bullet hd(s) \bullet z''}_{p \bullet s}, h^{z \bullet hd(s) \bullet z''}_{p' \bullet s'})$-good tuples $(z \cdot hd(s) \cdot z'' : (p \cdot s, \epsilon) : (p' \cdot s', \epsilon))$. Hence, it will also return $z \cdot z'$.

  ii) $j = h^{z \bullet z'}_{p \bullet s}(|p| + 1) = h^{z \bullet z'}_{p' \bullet s'}(|p'| + 1)$. This corresponds to the item (2) in Definition (1). Here $z_1 = (z \cdot z')_{|z|+1}$ is playing the role of $w_j$, $(p \cdot s)_{|p|+1} = hd(s)$ the one of $u_i$, $(p' \cdot s')_{|p'|+1} = hd(s')$ the one of $u'_{i'}$ and $\epsilon$ the one of $v$ and $v'$. As a result, the condition (2) of Definition (1) states that: both $lc(u_i) \sqcap_\mathbb{C} lc(u'_{i'})$ and $rc(u_i) \sqcap_\mathbb{C} rc(u'_{i'})$ are different from $\bot_\mathbb{C}$; and that $w_j = (lc(u_i) \sqcap_\mathbb{C} lc(u'_{i'}), (u_i)_{st}, rc(u_i) \sqcap_\mathbb{C} rc(u'_{i'}))$. This guarantees that the guard at line (8) will be satisfied, and that the computed $e$ at line (9) equals $w_j$. $\text{zip}(z : (p,s) : (p', s'))$ will then call $\text{zip}(z \cdot e : (p \cdot hd(s), tl(s)) : (p' \cdot hd(s'), tl(s')))$ which satisfies the induction hypothesis (since $|tl(s)| + |tl(s')| = n_0 - 1$) and returns all $(h^{z \bullet e \bullet z''}_{p \bullet s}, h^{z \bullet e \bullet z''}_{p' \bullet s'})$-good tuples $(z \cdot e \cdot z'' : (p \cdot s, \epsilon) : (p' \cdot s', \epsilon))$. Hence, it will also return $z \cdot z'$.

  iii) $j = h^{z \bullet z'}_{p' \bullet s'}(|p'| + 1)$ and $h^{z \bullet z'}_{p \bullet s}(|p|) < j < h^{z \bullet z'}_{p \bullet s}(|p| + 1)$. Symmetrical to the first case above, with lines (10) and (11) playing the role of lines (3) and (4).

The cases where one of $s, s'$ is empty are similar to taking one of (i) or (iii) in each of the ⊆ and ⊇ directions.

Lemma (9) uses the result of Lemma (11) in order to establish that the result of $\text{zip}(\epsilon : (\epsilon, \varphi) : (\epsilon, \varphi'))$ exactly captures $[\![\varphi]\!] \cap [\![\varphi']\!]$.

**Lemma 9 (intersection).** *Given $\varphi, \varphi'$ in $\mathbb{SCW}$, $\mathtt{zip}(\epsilon : (\epsilon, \varphi) : (\epsilon, \varphi'))$ returns a set $\{\varphi_1, \ldots \varphi_n\}$ s.t. $(\varphi \sqsubseteq_{\mathbb{CW}} \varphi_i), (\varphi' \sqsubseteq_{\mathbb{CW}} \varphi_i)$ for each $i \in \overline{n}$, and $\cup_{i \text{ in } \overline{n}} \llbracket \varphi_i \rrbracket = \llbracket \varphi \rrbracket \cap \llbracket \varphi' \rrbracket$.*

*Proof.* The tuple $(\epsilon : (\epsilon, \varphi) : (\epsilon, \varphi'))$ is good wrt. the empty injections $h_\epsilon^\epsilon, h_\epsilon^\epsilon$. Lemma (11) states $\mathtt{zip}(\epsilon : (\epsilon, \varphi) : (\epsilon, \varphi'))$ returns $I = \{z \mid (z : (\varphi, \epsilon) : (\varphi', \epsilon)) \text{ is } (h_\varphi^z, h_{\varphi'}^z)\text{-good}\}$. Observe that Definition (1) guarantees each $z$ satisfies $(\varphi \sqsubseteq^{(h_\varphi^z)} z)$ and $(\varphi' \sqsubseteq^{(h_{\varphi'}^z)} z)$. Hence, $\llbracket z \rrbracket \subseteq \llbracket \varphi \rrbracket \cap \llbracket \varphi' \rrbracket$. Moreover, assume a word $w = w_1 \cdots w_n$ in $Q^*$ verifies both $(w \models^h \varphi)$ and $(w \models^{h'} \varphi')$. We exhibit a counted word $z$ in $I$ such that $w \in \llbracket z \rrbracket$. We let $w' = w_{a_1} \cdots w_{a_m}$ be the word obtained by only keeping in $w$ those indices that belong to the union of the images of $h$ and $h'$. More precisely, for each $k : 1 \leq k \leq m$, we keep $w_{a_k}$ iff $a_k \in h(\overline{|\varphi|}) \cup h'(\overline{|\varphi'|})$. This defines an injection $h'' : \overline{|w'|} \to \overline{|w|}$ with $h''(k) = a_k$ for each $k \in \overline{|w'|}$. In the same context, we define the injections $(h_\varphi^{w'}) : \overline{|\varphi|} \to \overline{|w'|}$ and $(h_{\varphi'}^{w'}) : \overline{|\varphi'|} \to \overline{|w'|}$ with $(h_\varphi^{w'})(i) = k$ when $h(i) = a_k$ for each $i \in \overline{|\varphi|}$, and $(h_{\varphi'}^{w'})(i') = k$ when $h'(i') = a_k$ for each $i' \in \overline{|\varphi'|}$. Observe that by construction and well formedness of $\varphi$ and $\varphi'$, both $(w' \models^{(h_\varphi^{w'})} \varphi)$ and $(w' \models^{(h_{\varphi'}^{w'})} \varphi')$ hold. Back to Definition (1). Let $z$ be a counted word with $\underline{z} = w'$ and such that the tuple $(z : (\varphi, \epsilon) : (\varphi', \epsilon))$ is $(h_\varphi^{w'}, h_{\varphi'}^{w'})$-good. The word $z$ is well defined because $(w' \models^{(h_\varphi^{w'})} \varphi)$ and $(w' \models^{(h_{\varphi'}^{w'})} \varphi')$. Also, $w \models^{h''} z$ as otherwise, by construction of $z$, $w \not\models^h \varphi$ or $w \not\models^{h'} \varphi'$.

**Lemma 10 (separation).** *Assume $\mathtt{zip}(\epsilon : (\epsilon, \varphi) : (\epsilon, \varphi'))$ returns an empty set. If $\mathtt{zip}(\epsilon : (\epsilon, \nabla_\rho(\varphi)) : (\epsilon, \varphi) : (\epsilon, \varphi'))$ in Procedure augzip returns the pair $(\_, avoid)$, then any resolution $\rho'$ that satisfies $(avoid[\rho'])$ ensures $\mathtt{zip}(\epsilon : (\epsilon, \nabla_{\rho_{max}}(\varphi)) : (\epsilon, \varphi'))$ returns the empty set, with $\rho_{max}(q) = max(\rho(q), \rho'(q))$ for each $q$ in $Q$.*

*Proof.* Sketch. Suppose $(avoid[\rho'])$ holds but a counted word $z$ is still returned by $\mathtt{zip}(\epsilon : (\epsilon, \nabla_{\rho_{max}}(\varphi)) : (\epsilon, \varphi'))$. By construction of Procedure augzip, the formula $avoid$ has to imply the disjunction over the results of $(\mathtt{reasons}(\theta_i))$, where $\theta_i$ captures the sequence of tests that need to be validated in order to add $z$ to $\mathtt{zip}(\epsilon : (\epsilon, \nabla_\rho(\varphi)) : (\epsilon, \varphi'))$. Since $\rho'$ satisfies the disjunction, at least one of the tests will have to fail when trying to build $z$ in $\mathtt{zip}(\epsilon : (\epsilon, \nabla_{\rho_{max}}(\varphi)) : (\epsilon, \varphi'))$. Failure of the test is guaranteed since relaxation ensures that the precision of a counter in $\nabla_{\rho_{max}}(\varphi)$ is larger or equal (in the multiset sense) than the one of $\kappa(\nabla_{\rho'}(cr))$, which is already sufficient to fail at least one of the tests required to generate $z$ and captured by $\theta_i$.